\documentclass[conference]{IEEEtran}
\usepackage{amsmath,amssymb,amsfonts,amsthm}
\usepackage{algorithm,algorithmic,booktabs,bm,cite,graphicx,microtype,xcolor,float,scalerel}
\usepackage[caption=false,font=footnotesize]{subfig}

\newtheorem{lemma}{Lemma}

\newcommand{\E}[1]{\mathsf{E}\left[{#1}\right]}   

\newcommand{\mat}[1]{\mathsf{#1}}
\usepackage[hidelinks]{hyperref}
\usepackage{graphicx}
\usepackage{xcolor}

\newcommand{\vect}[1]{\mathbf{#1}} 
\newcommand{\vg}{\vect{g}}

\newcommand{\vN}{\vect{N}}
\newcommand{\vm}{\vect{m}}
\newcommand{\vx}{\vect{x}}
\newcommand{\vy}{\vect{y}}
\newcommand{\vY}{\vect{Y}}
\newcommand{\vs}{\vect{s}}
\newcommand{\vS}{\vect{S}}

\newcommand{\vZ}{\vect{Z}}

\newcommand{\vW}{\vect{W}}
\newcommand{\vh}{\vect{h}}
\newcommand{\vH}{\vect{H}}
\newcommand{\vu}{\vect{u}}
\newcommand{\vU}{\vect{U}}
\newcommand{\vL}{\vect{L}}
\newcommand{\vq}{\vect{q}}
\newcommand{\vdelta}{{\bm \delta}}
\newcommand{\vmu}{{\bm \mu}}

\newcommand{\vepsilon}{{\bm \epsilon}}
\newcommand{\vxi}{{\bm \xi}}
\newcommand{\vG}{\vect{G}}
\newcommand{\vzeros}{\vect{0}}

\newcommand{\vX}{\vect{X}}
\def\SS{\textrm{S}$\alpha$\textrm{S} } 
\def\SaS{\textrm{S}\alpha\textrm{S} } 
\def\rH{\rho_{\hspace{-0.5pt}{\scaleto{H}{4pt}}}}
\def\vmH{\vmu_{\hspace{-0.5pt}{\scaleto{H}{4pt}}}}
\def\gH{\gamma_{\hspace{-0.5pt}{\scaleto{H}{4pt}}}}

\begin{document}

\title{Temporal Diffusion Models for Joint Signal-Channel Estimation under Gaussian and $\alpha$-Stable MIMO Channels\thanks{This work was supported by the Maroun Semaan Faculty of Engineering and Architecture through the Cluster for Responsible Communication and Sensing.}}

\author{
\IEEEauthorblockN{Rawad Kansoh, Bahaa Ammoury, Charbel Hannoun, Jihad Fahs and Ibrahim Abou-Faycal}
\IEEEauthorblockA{
Department of Electrical and Computer Engineering\\
American University of Beirut, Beirut, Lebanon\\
Email: \{rak104, bna14, cnh10\}@mail.aub.edu,
\{jihad.fahs, ibrahim.abou-faycal\}@aub.edu.lb}}
\maketitle
\begin{abstract}
Diffusion models have emerged as powerful tools for learning complex data distributions and solving inverse problems. However, most diffusion models rely on the Gaussian distribution, which does not accurately represent heavy-tailed data. In this work, we study diffusion-based reconstruction under Gaussian and symmetric $\alpha$-stable fading channels, while also exploring the benefits of using possible temporal dependence in the channel. We consider two variants: pilot-based channel-only estimation and joint image-channel recovery for MIMO systems. Our approach is based on several standard and score-based diffusion models for both Gaussian and $\alpha$-stable distributions. We refine these models by guiding the reverse diffusion process using the observation likelihood and show how a single trained diffusion model can be reused for the location-scale family of distributions without retraining. In blind MIMO recovery, separate diffusion models are used for the image and channel, and their estimates are combined through likelihood guidance. For each setting, we compare independent and temporal reconstruction, where information from previous observations is used to improve the estimation. We show that, under a time-dependent channel model, such a temporal reconstruction improves channel estimation and the quality of the reconstructed image . Finally, we show that, for the same heavy-tailed channel data, using an $\alpha$-stable diffusion prior provides better channel estimation and image reconstruction than using a Gaussian diffusion prior.
\end{abstract}
\section{Introduction}
Reliable recovery of signals from noisy measurements is a central problem in communication systems and statistical signal processing. In an Integrated Sensing and Communication (ISAC) setting, the receiver must infer both the transmitted information and the unknown propagation effects, often from observations that are incomplete, distorted, or corrupted by noise. Classical Bayesian estimation provides a principled way to combine an observation model with prior information; its performance, however, depends on how faithfully the prior and likelihood represent the underlying physical process.

Diffusion generative models have emerged as flexible tools for learning complex data distributions and incorporating them into inverse problems. Denoising Diffusion Probabilistic Models (DDPM)s construct generative priors through a sequence of Gaussian perturbations and denoising transitions~\cite{ho2020ddpm}. Their continuous-time score-based counterparts describe related processes through stochastic differential equations~\cite{song2021scorebased}. For inverse problems, pretrained diffusion priors can be combined with observation likelihoods through guidance~\cite{chung2023dps}. Piecewise guidance has also been proposed to reduce the computational cost of posterior sampling~\cite{mohseni2026piecewise}. In communication systems, diffusion priors have been used to estimate Multiple-Input Multiple-Output (MIMO) channels from known pilots~\cite{arvinte2023mimo,fesl2024diffusion} and to jointly estimate the channel and detect unknown transmitted data~\cite{zilberstein2023joint}. More recently, separately-trained channel and source priors have been coupled through the observation model for completely pilot-free blind MIMO semantic communication~\cite{jiang2025blind}. These methods provide useful learned priors for reconstruction problems in which an explicit prior distribution is unavailable.

Two characteristics of communication systems motivate this work. First, observations are often collected sequentially, and signals and wireless channels may remain statistically dependent over consecutive time steps. Processing each observation independently therefore ignores potentially useful information from previous observations, whereas a temporal prior can exploit this dependence to improve reconstruction. Second, although Gaussian diffusion models are widely used, they provide poor coverage of heavy-tailed data~\cite{yoon2023lim,shariatian2025dlpm}. Alpha-stable distributions provide a natural extension to the heavy-tailed regime: like the Gaussian distribution, they arise as limits of normalized sums of independent random variables~\cite{kolmo1968}, with the Gaussian being a special case~\cite{samorodnitsky1994stable}. They therefore constitute a canonical model for heavy-tailed wireless fading~\cite{fahs2026}. In the sequel, \emph{stable} refers to both the Gaussian and \(\alpha\)-stable cases.

Generative diffusion formulations for heavy-tailed distributions have already been developed, including L\'{e}vy-It\^{o} Models (LIM) and Denoising L\'{e}vy Probabilistic Models (DLPM)~\cite{yoon2023lim,shariatian2025dlpm}. In this work, we investigate diffusion-based inference for Gaussian and symmetric $\alpha$-stable models in two settings: pilot-based channel reconstruction and blind image--channel recovery. We apply DDPM and score-based models for Gaussian distributions,
and DLPM and LIM for $\alpha$-stable distributions after implementing the necessary changes in terms of training loss, network parameterization, and reverse sampling procedure. Our results show that temporal diffusion priors improve reconstruction when the channel evolves according to a time-dependent model. We also demonstrate that a model trained on a standard distribution can be adapted without retraining to different locations, scales, and Gaussian means and covariances. Finally, we show that,  whenever the fading is heavy-tailed,  $\alpha$-stable diffusion priors provide better channel and image reconstruction than Gaussian diffusion priors.


\section{Background}

\subsection{Diffusion Models Conventions}

Throughout the paper, DDPM and score-based diffusion denote discrete and continuous-time Gaussian formulations of~\cite{ho2020ddpm,song2021scorebased}, based on noise prediction and score estimation, respectively. DLPM and LIM denote their discrete and continuous-time $\alpha$-stable counterparts
\cite{shariatian2025dlpm,yoon2023lim}, based on stable-noise prediction and
fractional-score estimation. A guided diffusion sample denotes the final output
of one guided reverse trajectory.

\subsection{Symmetric Alpha-Stable Distributions}
\label{sec:stable_background}

In this section, we define {\em stable\/} distributions and provide some useful properties. For more details, the reader is
referred to~\cite{samorodnitsky1994stable}. Stable Random Variables (RV)s satisfy
\begin{equation*}
aX_1+bX_2
\overset{\mathrm{d}}{=}
\sqrt[\alpha]{a^\alpha+b^\alpha} \, X + r,
\qquad \forall \, a,b > 0,
\end{equation*}
for some scalar $r$, where $X_1$ and $X_2$ are independent copies of $X$ and $\alpha \in (0,2]$ is the stability index. The case $\alpha = 2$ recovers the Gaussian law. In this work, we say RV $X$ is {\em Symmetric $\alpha$-Stable\/} (\SS\!\!) and write
$X\sim \SaS(\delta,\gamma)$, $\alpha \in (0,2)$, whenever
\begin{equation*}
\E{e^{iuX}} =
\exp\!\left(i\delta u-\gamma^\alpha|u|^\alpha\right), \qquad u \in \mathbb{R},
\end{equation*}
where $\delta\in\mathbb{R}$ and $\gamma>0$ are its {\em location\/} and {\em scale\/}, respectively. The index
$\alpha$ controls the tail asymptotics: a smaller $\alpha$ produces heavier
power-law tails of order $\sim \frac{1}{|x|^{1+\alpha}}$. Moreover, \(\E{|X|^p}<\infty\), whenever $p<\alpha$, so the variance is infinite for $\alpha<2$.

In addition, a random vector $\vX \in \mathbb{R}^{d}$ is said to be ``circular" \SS with law $\SaS(\vdelta, \gamma)$ if and only if
\begin{equation*}
\E{e^{i\vu^{\mathsf T}\vX}}
= \exp\!\left( i \vdelta^{\mathsf T}\vu
-\gamma^\alpha\|\vu\|_2^\alpha \right), \qquad \vu\in\mathbb{R}^{d}.
\end{equation*}
It admits the Gaussian scale-mixture representation
\begin{equation}
\vX
\overset{\mathrm{d}}{=}
\vdelta+\gamma\sqrt{A}\,\vG,
\label{eq:stable_subgaussian_representation}
\end{equation}
where
\(
\vG\sim\mathcal{N}(\vzeros,\mat{I}_d)
\)
is independent of $A$ --a positive $\tfrac{\alpha}{2}$-stable variable with scale equal to \(
\cos^{2/\alpha}\!\left(\frac{\pi\alpha}{4}\right).
\) Consequently, given that $A$ is equal to $a$
\begin{equation*}
\vX\mid (A=a) \;
\sim
\mathcal{N}\!\left(
\vdelta,
\gamma^2 a \, \mat{I}_d
\right).
\end{equation*}


\subsection{Reconstruction Metrics}
\label{subsec:reconstruction_metrics}

For $\vq_k,\widehat{\vq}_k\in\mathbb{R}^{d}$ we use the Normalized Mean Squared-Error
\begin{equation}
\operatorname{NMSE}_{\mathrm{dB}}
= 10 \log_{10}
\frac{ \sum_{k=1}^{K}
\left\| \widehat{\vq}_k-\vq_k
\right\|_2^2}{\sum_{k=1}^{K} \|\vq_k\|_2^2},
\label{eq:reconstruction_nmse}
\end{equation}
for Gaussian data, and
\begin{align}
\mathcal{E}_p &= \frac{1}{Kd}
\sum_{k=1}^{K} \sum_{i=1}^{d}
\left| \widehat q_{k,i}-q_{k,i}
\right|^p, \quad
0 < p < \alpha, \label{eq:reconstruction_fractional_error} \\
\& \quad \mathcal{E}_{\log,p} &= \frac{1}{Kd} \sum_{k=1}^{K}
\sum_{i=1}^{d} \log \!\left(
1 + \left| \widehat q_{k,i}-q_{k,i}
\right|^p \right),
\label{eq:reconstruction_log_fractional_error}
\end{align}
for $\alpha$-stable ones. Note that $\mathcal{E}_1$ is the mean absolute error. Image reconstruction is
evaluated using MS-SSIM~\cite{wang2003msssim}. Lower errors and higher MS-SSIM
indicate better performance.

\section{Problem Formulation}
\label{sec:problem_formulation}

With a view towards MIMO setups where images are transmitted using $N_t$ transmit antennas and $N_r$ receive antennas, define 
\begin{equation*}
\mat{X}_k =
\begin{bmatrix}
\vX_{k,1} & \cdots & \vX_{k,L}
\end{bmatrix}
\in\mathbb{R}^{N_t\times L},
\end{equation*}
where $\vX_{k,\ell} \in \mathbb{R}^{N_t}$ is the $\ell$-th transmitted symbol. We adopt a block-fading model where the channel matrix $\mat{H}_k \in \mathbb{R}^{N_r\times N_t}$ is assumed constant over these $L$
symbols. For each $\ell = 1, \cdots, L$, the received symbol is
\begin{equation*}
\vY_{k,\ell} = \mat{H}_k \vX_{k,\ell} + \vN_{k,\ell},
\qquad \vY_{k,\ell}, \vN_{k,\ell} \in \mathbb{R}^{N_r}.
\end{equation*} Stacking the $L$ received symbols we write
\begin{equation}
\mat{Y}_k = \mat{H}_k \mat{X}_k + \mat{N}_k, \qquad \mat{Y}_k, \mat{N}_k \in \mathbb{R}^{N_r\times L}.
\label{eq:joint_mimo_problem}
\end{equation}

From one block to another, we model the channel as a temporally-dependent MIMO channel that --while stationary, evolves temporally according to a first-order autoregressive model; it may follow either a Gaussian or a symmetric \(\alpha\)-stable law. For the Gaussian channel, the ``vectorized" channel matrix: \(\vH_k \hat{=} \operatorname{vec} (\mat{H}_k) \sim \mathcal{N}(\vmH, \mat{\Sigma}_H)\),
\begin{multline}
\vH_k = \vmH + \rH \left[ \vH_{k-1}-\vmH \right] + \sqrt{1-\rH^2} \, \vW_k \\ \text{where }\vW_k\sim\mathcal{N}(\vzeros,\mat{\Sigma}_H).
\label{eq:gaussian_channel_dynamics}
\end{multline}
For the symmetric \(\alpha\)-stable channel,
\begin{multline}
\vH_k = \vmH + \rH \left[\vH_{k-1} - \vmH \right] + \sqrt[\alpha]{ \left( 1-|\rH|^\alpha \right)} \, \vW_k \\
\text{where } \vW_k \sim \SaS
\!\left(\vzeros,\gH \right)
\label{eq:stable_channel_dynamics}
\end{multline}
preserving the stationary marginal \(\vH_k
\sim \SaS \left( \vmH, \gH \right)\). We restrict $\alpha > 1$ to guarantee a finite mean, in accordance with the LIM model~\cite{yoon2023lim}. 

The additive noise $\{\mat{N}_k\}_k$ in~\eqref{eq:joint_mimo_problem} are independent and chosen from the corresponding family: Gaussian noise for the Gaussian channel and \SS noise for the heavy-tailed channel. We study two inference scenarios that differ in what is known at the receiver and in the unknown quantities to be recovered.

\subsection{Channel-Only Estimation}
\label{subsec:problem_channel_only}

In the first scenario, a known pilot sequence is used and set, without loss of generality, to $\mat{X}_{\mathrm p} = \mat{I}$. The channel coefficients are processed jointly as a vector, giving the vectorized observation model
\begin{equation}
\vY_k=\vH_k+\vN_k.
\label{eq:channel_only_model}
\end{equation}
The objective is to estimate the temporally-varying channel vector $\vH_k$ from the pilot observations. This setting isolates
channel estimation from data detection and allows us to compare independent inference, which ignores past observations, with temporal inference, which uses
the previous channel estimate to form a predictive prior. All channel coefficients are reconstructed jointly.

\subsection{Joint Channel--Image MIMO Estimation}
\label{subsec:problem_joint_mimo}

In the second scenario, an initial training stage provides the channel $\mat{H}_0$. For $k>0$, no additional pilots are transmitted, and both the current channel and transmitted image are unknown. Let $\mat{D}_k$\ be an MNIST image \cite{lecun1998mnist}. A fixed encoder maps it to
\begin{equation*}
\mat{X}_k = f_\gamma(\mat{D}_k) =
\begin{bmatrix}
\vX_{k,1} & \cdots & \vX_{k,L}
\end{bmatrix}
\in \mathbb{R}^{N_t\times L},
\end{equation*}
where $\vX_{k,\ell} \in \mathbb{R}^{N_t}$ is the symbol vector transmitted
during channel use $\ell$. As previously stated, the channel $\mat{H}_k\in\mathbb{R}^{N_r\times N_t}$ is assumed constant over these $L$
channel uses. 

The receiver must jointly estimate \(\mat{H}_k\) and the information-bearing encoded signal \(\mat{X}_k = f_\gamma(\mat{D}_k)\), or equivalently the underlying image \(\mat{D}_k\). Separate learned priors are used for the channel and image, while the two unknowns are coupled during inference through the MIMO likelihood. The known initial channel \(\mat{H}_0\) anchors the temporal channel prior for subsequent time steps.

\section{Methodology}
\label{sec:methodology}

The two scenarios in Section~\ref{sec:problem_formulation} use the same inference principle. Diffusion models provide priors for the unknown quantities, the physical observation model supplies the likelihood used to guide the reverse process, and temporal information enters through a predictive channel prior. The diffusion networks are trained once on canonical distributions and adapted during inference without retraining when the location, scale, Gaussian mean, or Gaussian covariance of the prior changes.

\subsection{Canonical Gaussian Adaptation}
\label{subsec:canonical_gaussian_adaptation}

For each Gaussian formulation, one diffusion model is trained on the $d$-dimensional \( \vZ_0 \sim \mathcal{N}(\vzeros, \mat{I}_d) \). In DDPM, \( \vepsilon_\theta \) predicts the Gaussian perturbation, whereas in the continuous formulation, \( \vs_\theta \) predicts the score. 

Let $\vepsilon_t \sim \mathcal{N}(\vzeros,\mat{I}_d)$. The canonical forward marginals are given by $\vZ_t =
\sqrt{\bar{\alpha}_t}\,\vZ_0
+\sqrt{1-\bar{\alpha}_t}\,\vepsilon_t$ for DDPM and $\vZ(t) = m(t) \vZ_0 + \sigma(t) \vepsilon(t)$ for the score-based model. The same networks are adapted, without retraining, to \(\vX_0 \sim \mathcal{N}(\vmu,\mat{\Sigma})\) and the perturbed marginals are
\begin{align*}
\vX_t & \sim \mathcal{N}\!\left(
\sqrt{\bar{\alpha}_t}\,\vmu,
\mat{C}_t \right), &
\mat{C}_t & =
\bar{\alpha}_t\mat{\Sigma}
+(1-\bar{\alpha}_t)\mat{I}_d,\\
\vX(t) & \sim
\mathcal{N}\!\left(
m(t)\vmu, \mat{C}(t) \right),
&
\mat{C}(t) & = m^2(t)\mat{\Sigma}
+\sigma^2(t)\mat{I}_d.
\end{align*}

At every reverse step, let
\begin{equation*}
\mat{C}_t=\mat{R}_t\mat{R}_t^{\mathsf T},
\qquad
\mat{C}(t)=\mat{R}(t)\mat{R}^{\mathsf T}(t)
\end{equation*}
The canonical inputs are
\begin{align*}
\vU_t & = \mat{R}_t^{-1} \left( \vX_t - \sqrt{ \bar{\alpha}_t} \, \vmu \right),
&& \text{(DDPM)},\\
\vU(t) & = \mat{R}^{-1}(t)
\left( \vX(t) - m(t)\vmu \right),
&& \text{(score-based)},
\end{align*}
and the network outputs are transported as
\begin{align}
\widehat{\vepsilon}_{\vmu,\mat{\Sigma}}(\vX_t,t)
&= \mat{R}_t^{-\mathsf T}
\vepsilon_\theta(\vU_t,t),
&& \text{(DDPM)}, \label{eq:ddpm_gaussian_adaptation}\\
\widehat{\vs}_{\vmu,\mat{\Sigma}}(\vX(t),t) & =
\mat{R}^{-\mathsf T}(t)
\, \vs_\theta(\vU(t),t),
&& \text{(score-based)}. \label{eq:score_gaussian_adaptation}
\end{align}
Note that the DDPM transport is stated for the MSE-optimal noise predictor. It is established using the joint Gaussianity of $(\vepsilon_t,\vX_t)$ and the fact that the MSE-optimal noise predictor is $\E{\vepsilon_t \mid \vX_t}$. Applying the Gaussian conditional-mean formula together with $\mat{C}_t^{-1}=\mat{R}_t^{-\mathsf T}\mat{R}_t^{-1}$ gives the stated input and output transformations (see Appendix~\ref{app:ddpm-comprehensive}). For the score-based formulation, the transport follows directly from the chain rule under the affine transformation $\vU(t) = \mat{R}^{-1}(t)(\vX(t) -m(t) \vmu)$ (see Appendix~\ref{app:score-comprehensive}).

\subsection{Canonical Alpha-Stable Adaptation}
\label{subsec:canonical_stable_adaptation}

For each $\alpha$-stable formulation, one diffusion model is trained on $\vZ_0 \sim \SaS (\vzeros,1)$. In DLPM, \(\vepsilon_\theta\) predicts the stable perturbation, whereas in LIM, \(\vs_\theta^{(\alpha)}\) predicts the fractional score defined in~\cite{yoon2023lim,fahs2026linearity}. We use the logarithmic error
$\mathcal{E}_{\log,p}$ defined in
\eqref{eq:reconstruction_log_fractional_error} in the noise-prediction models, with $p=1$ in the experiments.

The trained models are adapted without retraining to
\begin{equation*}
\vX_0 = \vmu + \gamma \, \vZ_0 \sim \SaS (\vmu, \gamma).
\end{equation*}
For DLPM, whose canonical forward marginal is \(\vZ_t = \bar c_t \vZ_0 + \left( 1-\bar c_t^\alpha\right)^{1/\alpha}{\vepsilon_t}\), {where $\vepsilon _t\sim
\SaS
\!\left(\vzeros,1 \right)$}, the experiments were conducted with \(\gamma=1\), and
\begin{align*}
\vU_t & = \vX_t - \bar c_t\vmu,
&
\widehat{\vepsilon}_{\vmu,\gamma}(\vX_t,t) & = \vepsilon_\theta(\vU_t,t).
\end{align*}

For LIM, whose canonical forward marginal is
\(\vZ(t) = a(t) \vZ_0 + \sigma(t) \vepsilon(t)\),
\begin{equation*}
\vU(t) = \frac{\vX(t)-a(t)\vmu}{\gamma},\,\,\,\,\,
 \vs_{\vmu,\gamma}^{(\alpha)}(\vX(t),t)
= \gamma^{1-\alpha}
\vs_\theta^{(\alpha)}(\vU(t),t).
\end{equation*}
For DLPM with $\gamma=1$, the transport follows directly by substituting $\vX_0 = \vmu + \vZ_0$ into the forward process (see Appendix~\ref{app:dlpm-comprehensive}). For LIM, the transport rule follows by applying the Fourier transform to the fractional score and examining its behavior under the location-scale transformation $\vX(t) = a(t) \vmu + \gamma \vU(t)$, hence the factor $\gamma^{1-\alpha}$ (see Appendix~\ref{app:lim-comprehensive}).

\subsection{Observation Conditioning and Joint Guidance}
\label{subsec:gaussian_observation_conditioning}
\label{subsec:stable_observation_conditioning}

For both scenarios, the diffusion priors are conditioned on the physical observation model. We use the generic vectorized representation
\begin{equation}
\vY = f\!\left( \vxi_0^{(1)}, \cdots, \vxi_0^{(J)} \right) + \vN,
\label{eq:generic_observation_model}
\end{equation}
where each unknown \(\vxi_0^{(j)}\) has a separately trained diffusion prior and reverse state \(\vxi_t^{(j)}\). Matrix-valued observations are understood through vectorization.

For Gaussian noise \( \vN \sim \mathcal{N}(\vzeros,\sigma_n^2\mat{I}_d\)), the reverse models provide clean estimates
\(\widehat{\vxi}_0^{(j)}(\vxi_t^{(j)},t)\), and the likelihood gradient for the \(j\)-th unknown is
\begin{equation*}
\vg_t^{(j)} = \nabla_{\vxi_t^{(j)}} \log p \!\left(
\vy \mid \widehat{\vxi}_0^{(1)}, \cdots, \widehat{\vxi}_0^{(J)}
\right).
\end{equation*}
A weighted version of \(\vg_t^{(j)}\) is added to the corresponding reverse update. Thus, separately trained priors are coupled only through the shared likelihood.

When the Gaussian observation is linear,
\begin{equation}
\vY = \mat{A} \vxi_0 + \vN, \qquad
\vxi_0 \sim \mathcal{N}(\vmu,\mat{\Sigma}),
\label{eq:linear_gaussian_model}
\end{equation}
the posterior is Gaussian with
\begin{align}
\vmu_{\mathrm{post}}
&=
\vmu
+\mat{\Sigma}\mat{A}^{\mathsf T}
\left(
\mat{A}\mat{\Sigma}\mat{A}^{\mathsf T}+\sigma_n^2\mat{I}_d\
\right)^{-1}
(\vy-\mat{A}\vmu),\label{eq:linear_gaussian_mean}\\
\mat{\Sigma}_{\mathrm{post}}
&=
\mat{\Sigma}
-\mat{\Sigma}\mat{A}^{\mathsf T}
\left(
\mat{A}\mat{\Sigma}\mat{A}^{\mathsf T}+\sigma_n^2\mat{I}_d\
\right)^{-1}
\mat{A}\mat{\Sigma}.
\label{eq:linear_gaussian_covariance}
\end{align}
This permits exact Gaussian conditional sampling in the channel-only setting.

For \SS noise \(\vN \sim \SaS (\vzeros, \gamma_n)\), we use the Gaussian scale-mixture representation in
\eqref{eq:stable_subgaussian_representation}. Conditioned on the positive-stable variable \(A\),
\begin{equation*}
\vY \mid \vxi_0^{(1)},\cdots,\vxi_0^{(J)}, A \quad 
\sim \mathcal{N}\!\left(
f\!\left(\vxi_0^{(1)},\cdots,\vxi_0^{(J)}\right),
\gamma_n^2 A \, \mat{I}_d
\right).
\end{equation*}
Using \(M_{\mathrm{MC}}\) samples
\(\{A^{(m)}\}_{m=1}^{M_{\mathrm{MC}}}\), the marginal likelihood is approximated by
\begin{multline*}
\widehat p\!\left(
\vy \mid \vxi_0^{(1)}, \cdots, \vxi_0^{(J)}
\right) \\ = \frac{1}{M_{\mathrm{MC}}} \sum_{m=1}^{M_{\mathrm{MC}}}
\mathcal{N}\!\left( \vy;
f\!\left( \vxi_0^{(1)}, \cdots, \vxi_0^{(J)} \right),
\gamma_n^2A^{(m)}\mat{I}_d
\right).
\end{multline*}
The gradient of the logarithm of this approximation is used in the same guided reverse-update construction.

\subsection{Temporal Channel Priors}
\label{subsec:temporal_priors}

Independent inference uses the stationary channel law at every time step. Temporal inference instead uses the previous reconstruction to form a predictive prior for the current channel. The specific uncertainty update depends on whether the preceding channel uncertainty is propagated explicitly or the previous estimate is treated as exact.

For the Gaussian channel, when the preceding posterior covariance is propagated,
\begin{align}
\vH_k \mid \widehat{\vh}_{k-1}
& \sim
\mathcal{N}\!\left(
\vm_{H,k}^{-},
\mat{P}_{H,k}^{-}
\right),
\label{eq:gaussian_predictive_channel_prior} \\
\text{where } \quad \vm_{H,k}^{-}
&=
\vmH+\rH(\widehat{\vh}_{k-1}-\vmH), \notag \\
\mat{P}_{H,k}^{-}
&=
\rH^2\mat{P}_{H,k-1}
+(1-\rH^2)\mat{\Sigma}_H. \notag
\end{align}
If \(\widehat{\vh}_{k-1}\) is treated as exact, this reduces to
\begin{equation}
\vH_k \mid \widehat{\vh}_{k-1}
\!\sim\!
\mathcal{N}\!\left(\!
\vmH+\rH(\widehat{\vh}_{k-1}-\vmH\!),
(1-\rH^2)\mat{\Sigma}_H\!
\right).
\label{eq:gaussian_mimo_temporal_prior}
\end{equation}

For the \SS channel, uncertainty propagation is {\em approximated\/} by
\begin{align}
\vH_k \mid\widehat{\vh}_{k-1} & \approx \SaS\!\left( \vmu_{H,k}^{-}, \gamma_{H,k}^{-}
\right),
\label{eq:stable_predictive_channel_prior} \\
\text{where } \quad
\vmu_{H,k}^{-} & = \vmH+\rH(\widehat{\vh}_{k-1}-\vmH), \notag \\
\gamma_{H,k}^{-} & = \sqrt[\alpha]{\left| \rH \right|^\alpha \widehat{\gamma}_{H,k-1}^{\alpha}
+ \left( 1-|\rH|^\alpha \right) \gH^\alpha }. \notag
\end{align}
This is an approximation because the distribution of the previous estimation error is generally not $\alpha$-stable. We therefore match a zero-location \SS approximation to the empirical interquartile range of the guided diffusion samples~\cite{mcculloch1986simple}. 

If the previous estimate is treated as exact, the predictive scale reduces to \( \gH \sqrt[\alpha]{1-|\rH|^\alpha} \), yielding
\begin{equation}
\vH_k \mid \widehat{\vh}_{k-1} \sim \SaS \! \left(
\vmH + \rH(\widehat{\vh}_{k-1} - \vmH),
\gH \!\sqrt[\alpha]{1\!-\!|\rH|^\alpha} \right).
\label{eq:mimo_stable_temporal_prior}
\end{equation}

\section{Experiments and Results}
\label{sec:results}

\subsection{Channel-Only Estimation}
\label{subsec:experiments_channel_only}

Using pilot signals as in \eqref{eq:channel_only_model}, we jointly process three channel coefficients, with $\vH_k\in\mathbb{R}^{3}$.
\subsubsection{Gaussian Channel}
\label{subsec:temporal_gaussian_reconstruction}
\label{subsec:results_temporal_gaussian}

As an example, we consider channels' coefficients with  correlated stationary distribution
\begin{equation*}
\vH_k \sim
\mathcal{N}\!\left(
\begin{bmatrix}
5 & 3 & -2
\end{bmatrix}^{\mathsf T},
\begin{bmatrix}
2 & 0.4 & 0.2\\
0.4 & 0.5 & 0.1\\
0.2 & 0.1 & 1.5
\end{bmatrix}
\right),
\end{equation*}
and \(\vN_k\sim\mathcal{N}(\vzeros,\mat{I}_3)\).
The independent method uses the stationary prior, whereas the temporal methods use the predictive prior in \eqref{eq:gaussian_predictive_channel_prior}. At \(k=0\), \(\mat{P}_{H,0}
=
\left(
\mat{\Sigma}_H^{-1}+\mat{I}_3
\right)^{-1}\), and the posterior covariance is propagated recursively for \(k>0\).

We compare  four estimators. \emph{Independent guided} and \emph{temporal guided} estimators use likelihood guidance, while \emph{independent conditional} and \emph{temporal conditional} estimators use exact Gaussian conditioning, both with stationary channel prior~\eqref{eq:gaussian_channel_dynamics} and predictive channel prior~\eqref{eq:gaussian_predictive_channel_prior}.

From the $P$ diffusion samples
$\{\vh_k^{(j)}\}_{j=1}^{P}$, the guided estimate is selected as
\begin{equation*}
\widehat{\vh}_k
=
\vh_k^{(j_k^\star)},
\qquad
j_k^\star
=
\arg\min_{1\leq j\leq M}
\left\|
\vy_k-\vh_k^{(j)}
\right\|^2.
\end{equation*}
The MSE-optimal estimator is the exact Bayesian posterior mean.
Performance is evaluated using NMSE defined in \eqref{eq:reconstruction_nmse}.

\begin{figure}[H]
\centering

\includegraphics[
    width=0.95\linewidth,
    keepaspectratio]{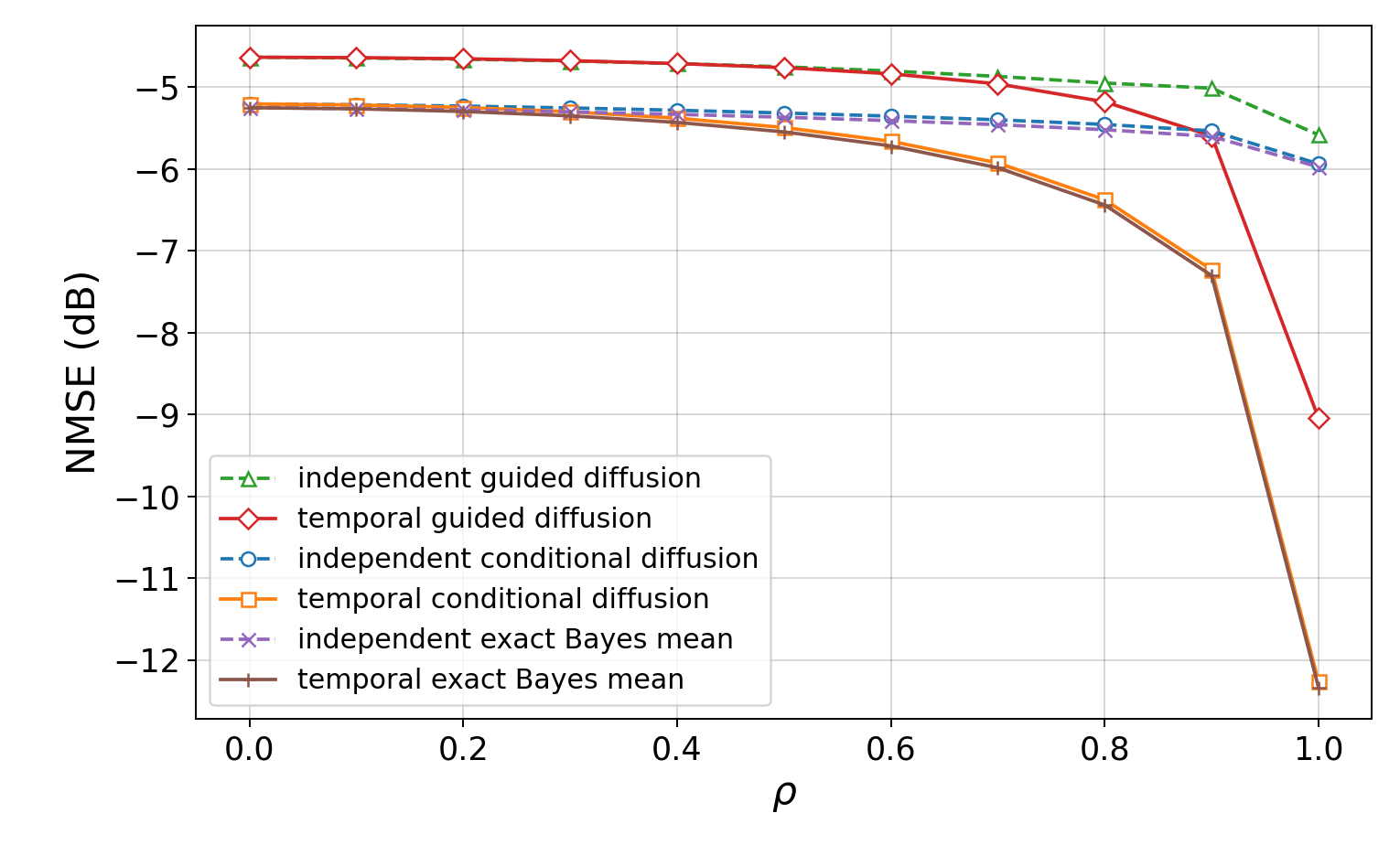}


\includegraphics[
    width=0.95\linewidth,
    keepaspectratio
]{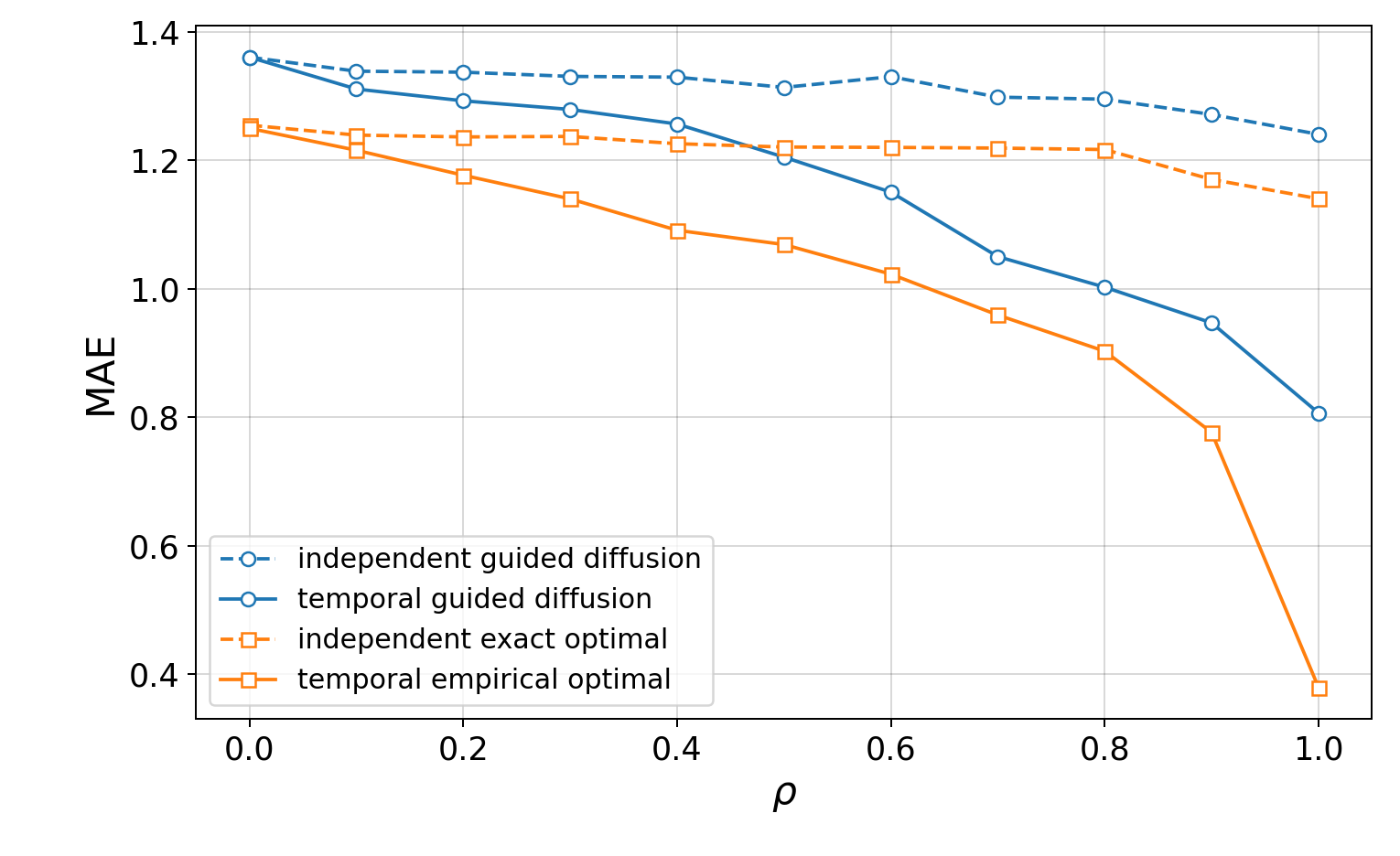}

\caption{Reconstruction loss versus \(\rH\) for independent and temporal channel reconstruction. Top: Gaussian channels using DDPM. Bottom: $\alpha$-stable channels using DLPM with \(\alpha=1.7\).}
\label{fig:temporal_rho_sweep}
\vspace{-4mm}
\end{figure}


\begin{figure*}[!t]
    \centering
    \includegraphics[
        width=0.485\textwidth
    ]{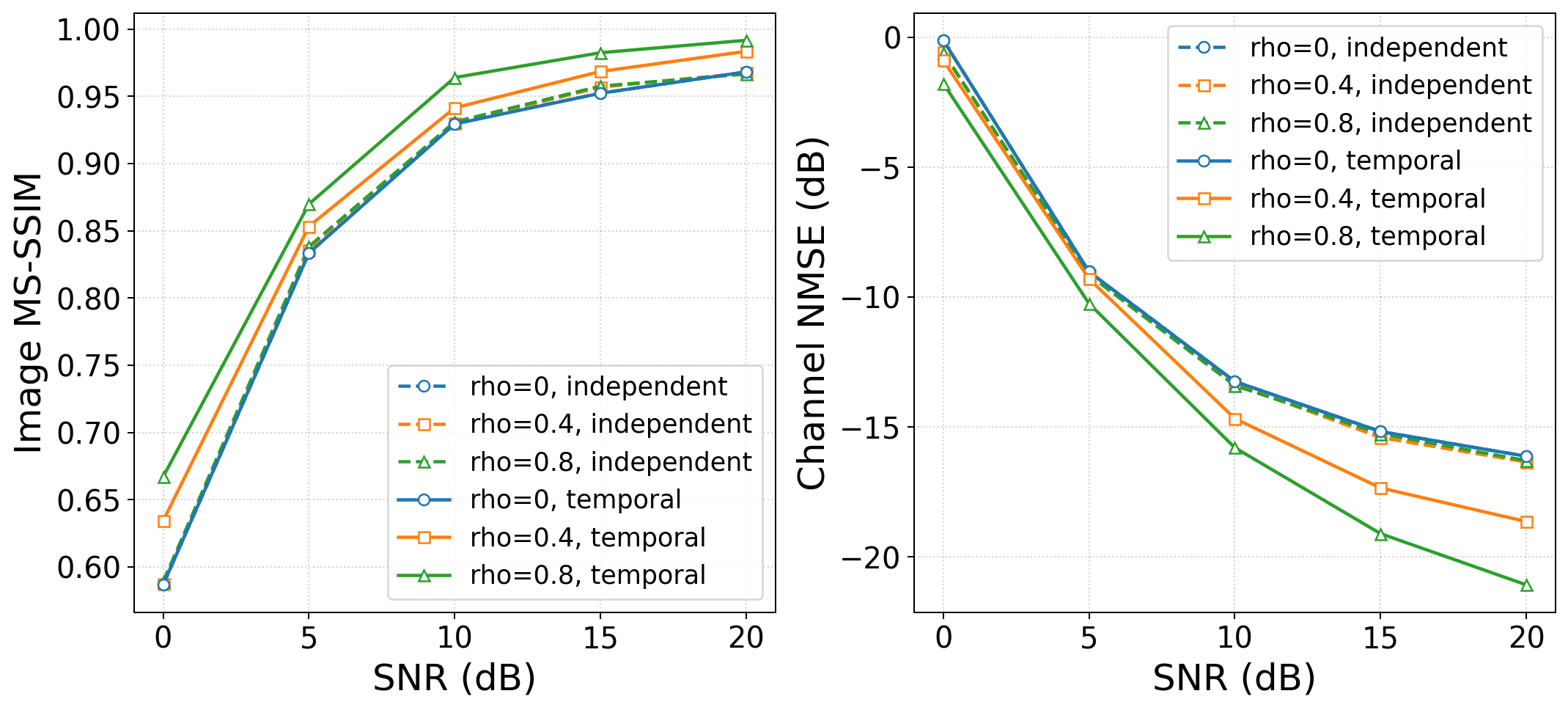}
    \hfill
    \includegraphics[
        width=0.485\textwidth
    ]{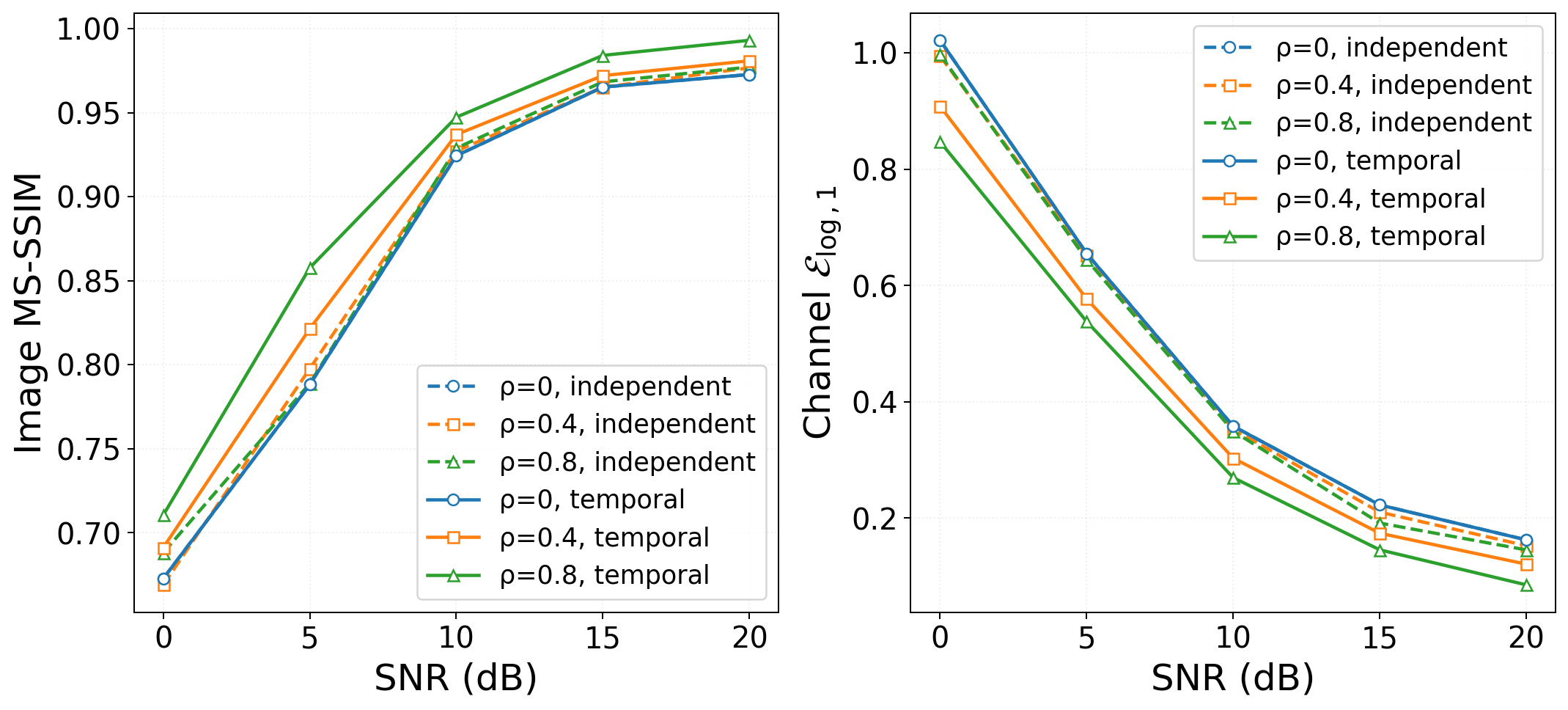}
    \caption{Image MS-SSIM and channel loss vs. SNR for independent and temporal joint channel--image MIMO reconstruction. Left: Gaussian channels using DDPM. Right: $\alpha$-stable channels with $\alpha=1.7$ using DLPM.}
    \label{fig:mimo_snr_results}
    \vspace{-3mm}
\end{figure*}

\subsubsection{Alpha-Stable Channel}
\label{subsec:temporal_stable_reconstruction}
\label{subsec:results_temporal_stable}

 We consider the following example for the \SS $\{\vH_k\}$s: stability index \(\alpha=1.7\), location \(\vmH=
\begin{bmatrix}
5 & 3 & -2
\end{bmatrix}^{\mathsf T}\),
and unit stationary scale. The additive $\alpha$-stable noise is \SS \(\vN_k \sim \SaS (\vzeros, 1)\). These $\alpha$-stable vectors can be generated using appropriate python packages such as SciPy or the AUB-HTP package~\cite{aub-htp}.

Independent inference uses the stationary channel prior~\eqref{eq:stable_channel_dynamics}, whereas temporal inference uses \eqref{eq:stable_predictive_channel_prior}. The channel estimate is the componentwise median of the final guided samples,
\begin{equation*}
\widehat{\vh}_k = \operatorname{median}
\left\{ \vh_k^{(1)}, \cdots, \vh_k^{(P)}
\right\}.
\end{equation*}

For equal signal and noise stability indices and scales, the exact independent absolute-error estimator is
\begin{equation*}
\widehat{\vh}_k^{\mathrm{ind,opt}}
=
\frac{\vy_k+\vmH}{2}.
\end{equation*}

For the temporal reference, the MAE-optimal estimator is the componentwise median of \( p( \vH_k \mid \vy_0, \vy_1, \cdots, \vy_k)\), approximated numerically using samples from the known temporal model weighted by the observation likelihood~\cite{doucet2001sequential}. Performance is evaluated using \(\mathcal{E}_1\) defined in
\eqref{eq:reconstruction_fractional_error}.

 \subsubsection*{Results} 
\begin{figure}[ht!]
\centering
\includegraphics[
    width=0.95\linewidth,
    keepaspectratio
]
{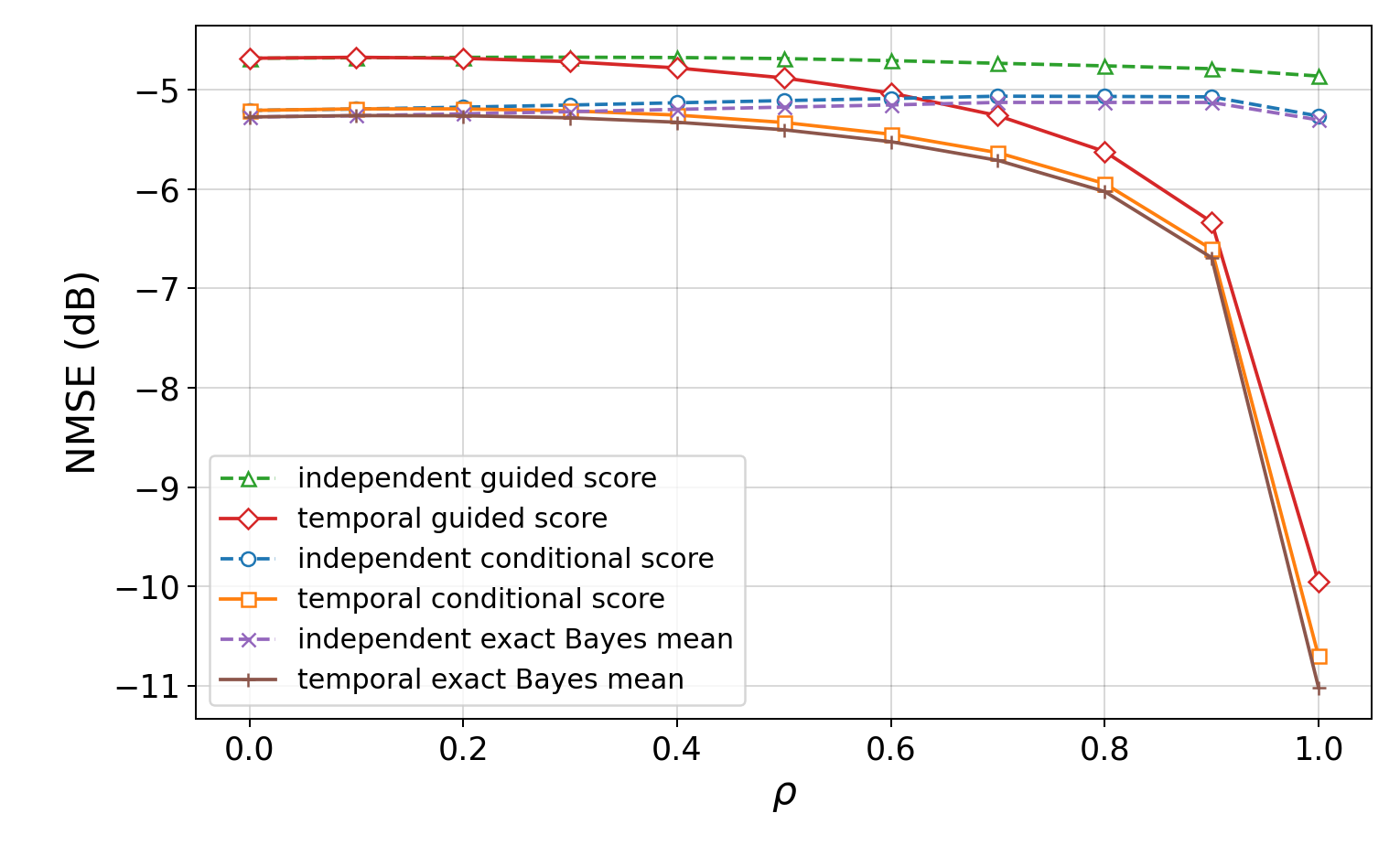}

\includegraphics[
    width=0.95\linewidth,
    keepaspectratio
]
{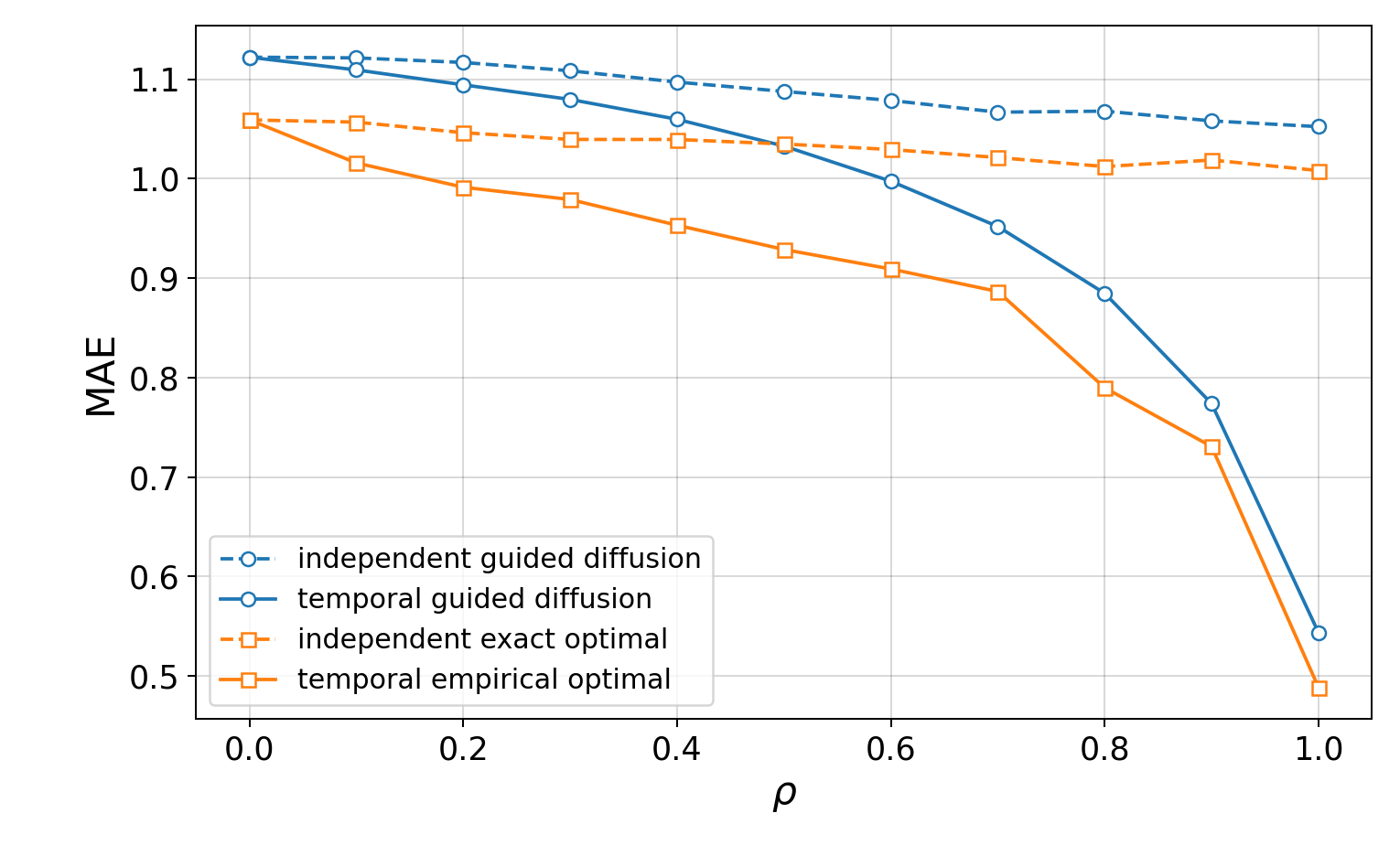}
\caption{Reconstruction loss versus \(\rH\) for independent and temporal
channel reconstruction. Top: Gaussian channels using the score-based model.
Bottom: $\alpha$-stable channels using LIM with \(\alpha=1.7\). This
corresponds to the channel-only experiment in
Section~\ref{subsec:experiments_channel_only}.}
\label{fig:appendix_continuous_channel_only}
\end{figure}

We use \(K=20\) sequences. Figure~\ref{fig:temporal_rho_sweep} shows that the gain of temporal over independent reconstruction increases with \(\rH\), measured by a reduction in NMSE for Gaussian channels with DDPM and in MAE for \(\alpha\)-stable channels with DLPM. For the Gaussian case, this behavior is observed under both guided and conditional reconstruction. At \(\rH=0\), the temporal and independent methods coincide, as expected. Similar trends are observed for the continuous score-based and LIM models in Fig.~\ref{fig:appendix_continuous_channel_only}.

\subsection{Joint Channel--Image MIMO Estimation}
\label{subsec:experiments_joint_mimo}

In this setup, each sequence contains encoded MNIST images represented as \(4\times32\) symbols $\{\mat{X}_k\}$ in~\eqref{eq:joint_mimo_problem}. The channel $\{ \mat{H}_k\}_k$ is \(4\times4\) with an "initial" matrix \(\mat{H}_0\) that is known from the training stage; for \(k>0\), the receiver jointly estimates the current channel and image. The image and channel priors are trained separately and coupled only through the common MIMO likelihood.

In the generic guidance model \eqref{eq:generic_observation_model}, the two unknowns are  \(\vxi_k^{(1)}=\operatorname{vec}(\mat{H}_k)\) and \(\vxi_k^{(2)}=\operatorname{vec}(\mat{D}_k)\), with $f(\mat{H}_k,\mat{D}_k) = \mat{H}_kf_\gamma(\mat{D}_k) = \mat{H}_k \, \mat{X}_k$.

For each channel family, we compare independent joint inference using the stationary channel prior with temporal joint inference using the corresponding predictive prior.

\subsubsection{Gaussian Channel}
\label{subsec:temporal_gaussian_mimo}
\label{subsec:results_temporal_gaussian_mimo}
The vectorized channel satisfies \(\vH_k=\operatorname{vec}(\mat{H}_k)
\sim
\mathcal{N}(\vzeros,\mat{I}_{16})\) and 
and the entries of \(\mat{N}_k\) are independent
\(\mathcal{N}(0,\sigma_N^2)\) variables. The temporal method uses \eqref{eq:gaussian_mimo_temporal_prior}, treating the previous channel estimate as exact, while the independent method uses the stationary channel prior~\eqref{eq:gaussian_channel_dynamics}. From the joint candidates \( \left\{ (\mat{H}_k^{(j)},\mat{D}_k^{(j)}) \right\}_{j=1}^{M} \), we select
\begin{equation*}
j_k^\star
=
\arg\min_j
\left\|
\mat{Y}_k
-\mat{H}_k^{(j)}f_\gamma(\mat{D}_k^{(j)})
\right\|_{\textrm{Frob}}^2.
\end{equation*}

\subsubsection{Alpha-Stable Channel}
\label{subsec:temporal_stable_mimo}
\label{subsec:results_stable_mimo}

The channel models are trained on the 16-dimensional \( \vH_k = \operatorname{vec}(\mat{H}_k) \sim \SaS (\vzeros, 1)\), for $\alpha=1.7$ and
the additive $\alpha$-stable noise is \SS \(\vN_k \sim \SaS (\vzeros, 1)\). Since $\alpha$-stable noise has infinite variance, its scale is selected according to
\begin{equation*}
\gamma_N
=
\frac{
\operatorname{RMS}\!\left(
\mat{H}_k f_\gamma(\mat{D}_k)
\right)
}{
10^{\mathrm{SNR}/20}
},
\qquad
\operatorname{RMS}(\mat{Z})
=
\frac{\|\mat{Z}\|_{\textrm{Frob}}}{\sqrt{N_rL}}.
\end{equation*}
The independent method uses the stationary channel prior~\eqref{eq:stable_channel_dynamics}, while the temporal method uses \eqref{eq:mimo_stable_temporal_prior}. From the joint candidates, we select the one with the largest approximated $\alpha$-stable measurement likelihood,
\[ \displaystyle
j_k^\star
=
\arg\max_{1\leq j\leq M} \; 
\widehat p\!\left(
\mat{Y}_k
\mid
\mat{H}_k^{(j)},\mat{D}_k^{(j)}
\right).
\]

\subsubsection*{Independent and temporal reconstruction results}
\begin{figure}[!t]
    \centering

    \includegraphics[
        width=0.95\linewidth,
        keepaspectratio
    ]{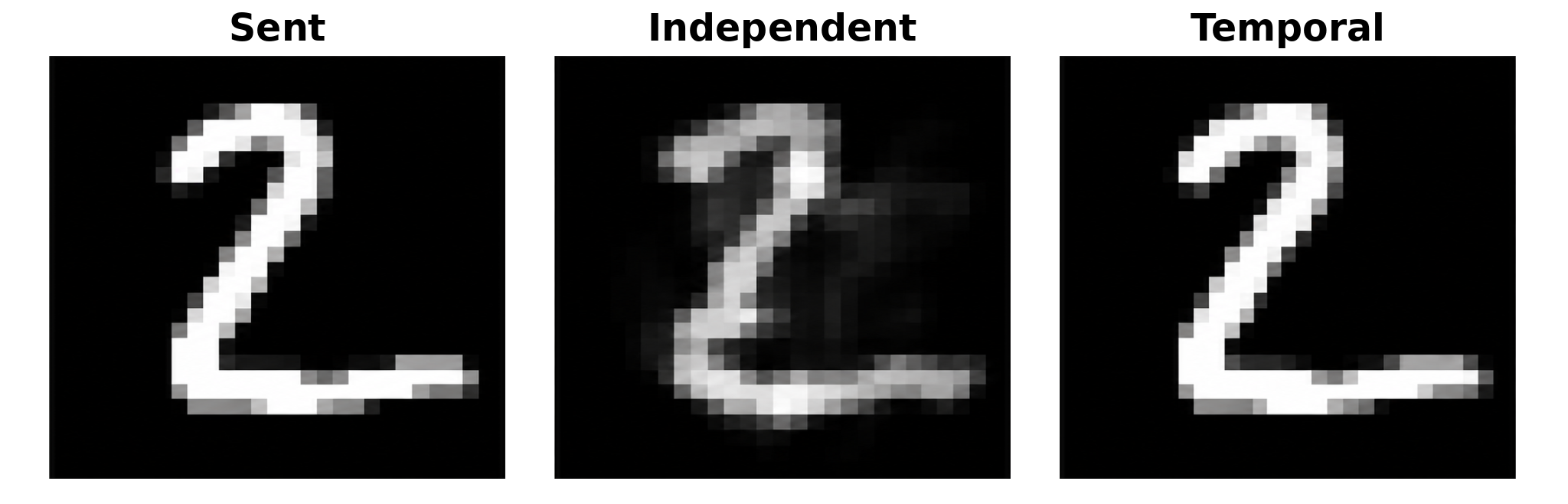}


    \includegraphics[
        width=0.92\linewidth,
        keepaspectratio
    ]{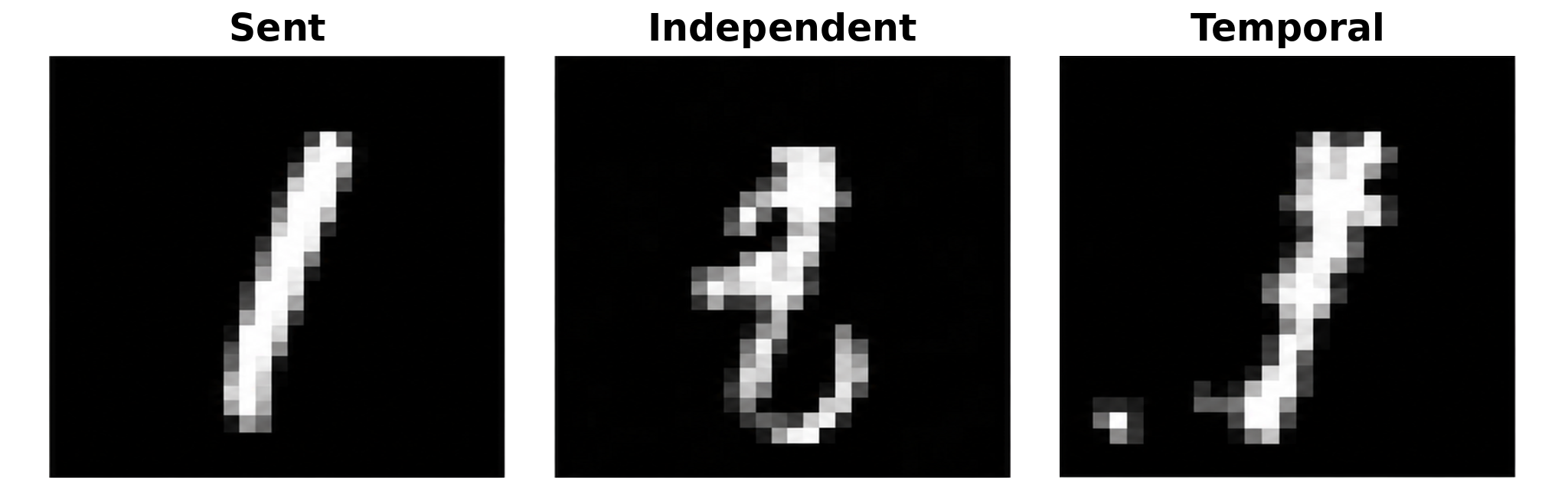}

    \caption{Sent images and the corresponding independent and temporal
    joint MIMO reconstructions for \(\rH=0.8\).
    Top: Gaussian channel using DDPM at SNR \(=10\) dB.
    Bottom: \(\alpha\)-stable channel using DLPM with
    \(\alpha=1.7\) at SNR \(=5\) dB.}
    \label{fig:mimo_reconstructed_images}
    \vspace{1mm}
\end{figure}

\begin{figure}[ht!]
    \centering
    \includegraphics[
        width=\linewidth,
        keepaspectratio
    ]{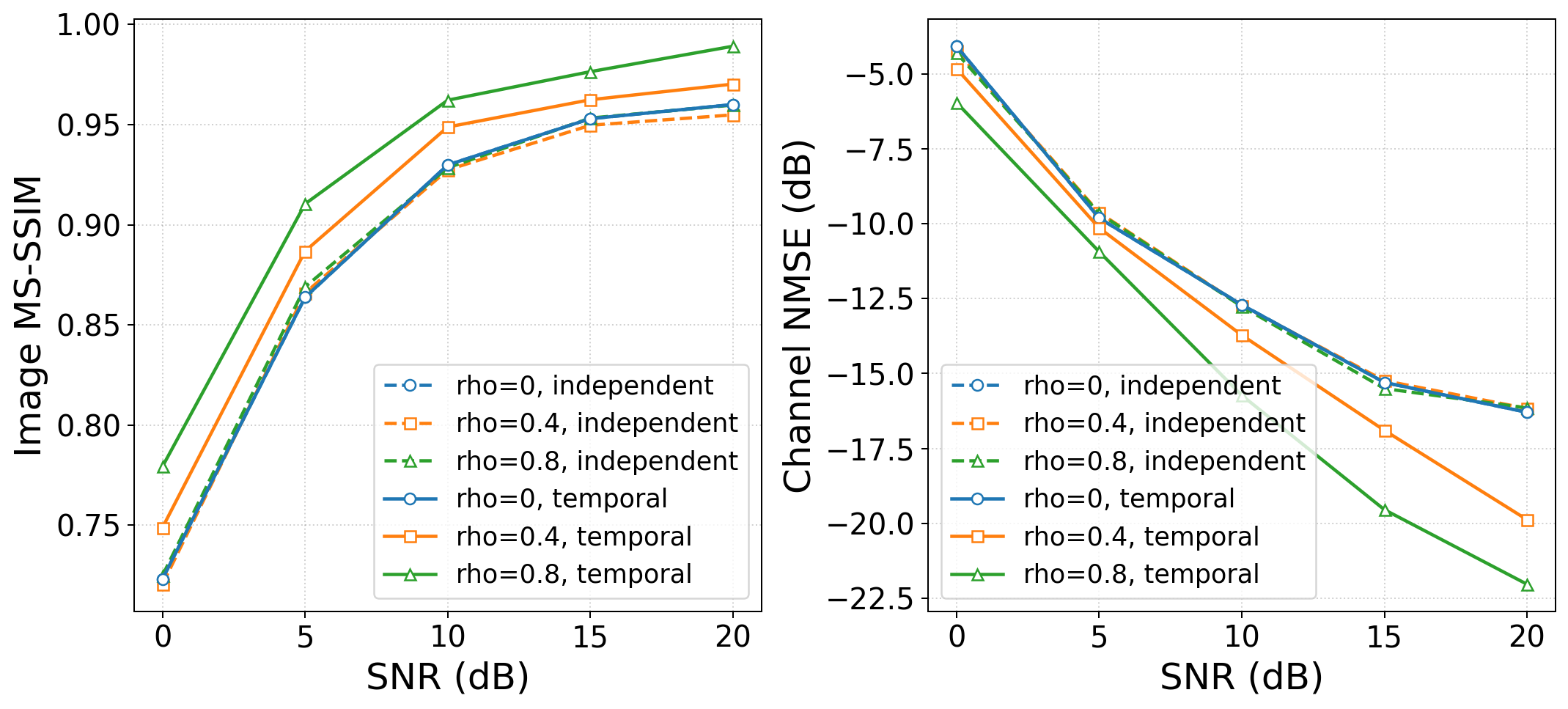}

    \par\smallskip

    \includegraphics[
        width=\linewidth,
        keepaspectratio
    ]{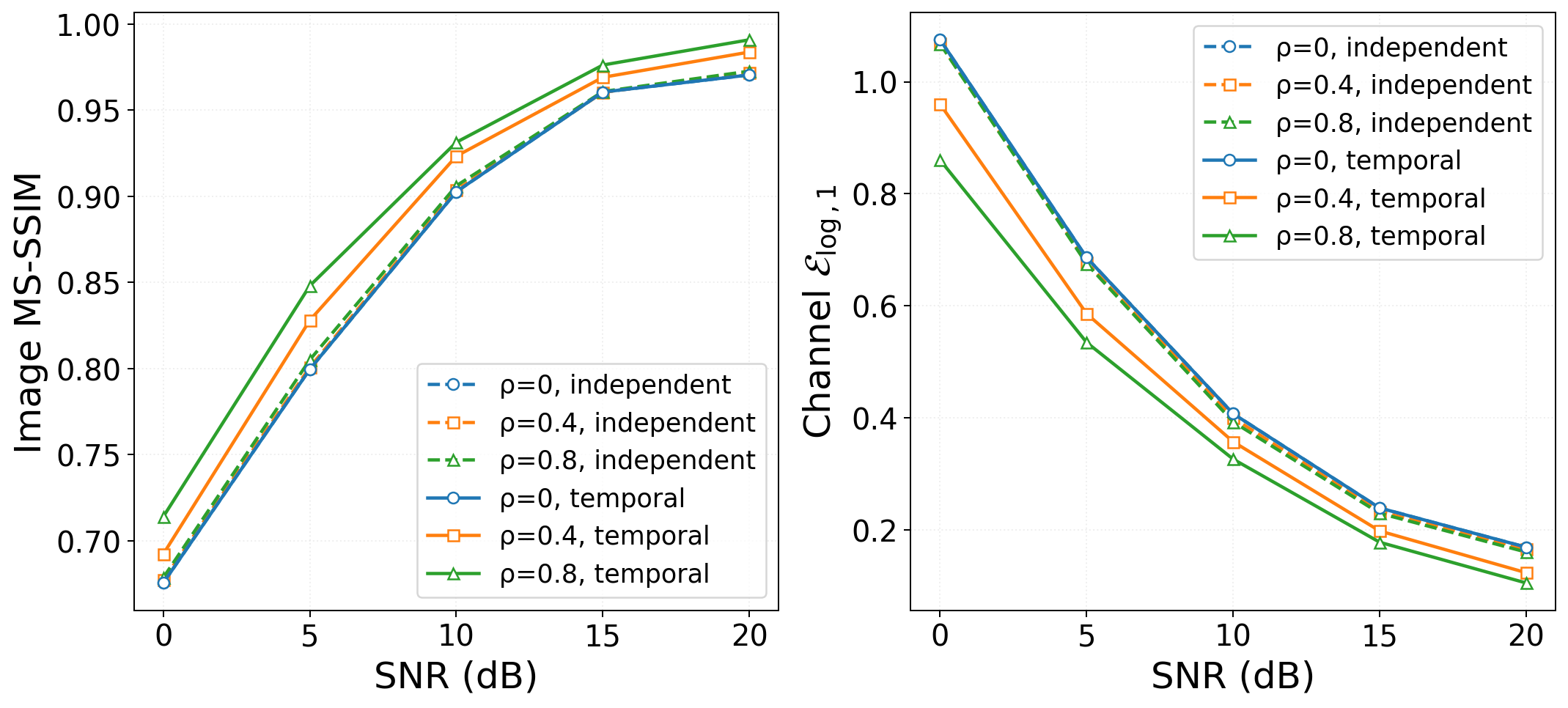}

    \caption{Image MS-SSIM and channel loss vs. SNR for independent and temporal
    joint channel--image MIMO reconstruction. Top: Gaussian channels using the
    score-based model. Bottom: $\alpha$-stable channels with $\alpha=1.7$ using
    LIM. This corresponds to the joint channel--image MIMO experiment in
    Section~\ref{subsec:experiments_joint_mimo}.}
    \label{fig:appendix_continuous_mimo}
\end{figure}

\begin{figure}[!ht]
    \centering

    \includegraphics[
        width=0.95\linewidth,
        keepaspectratio
    ]{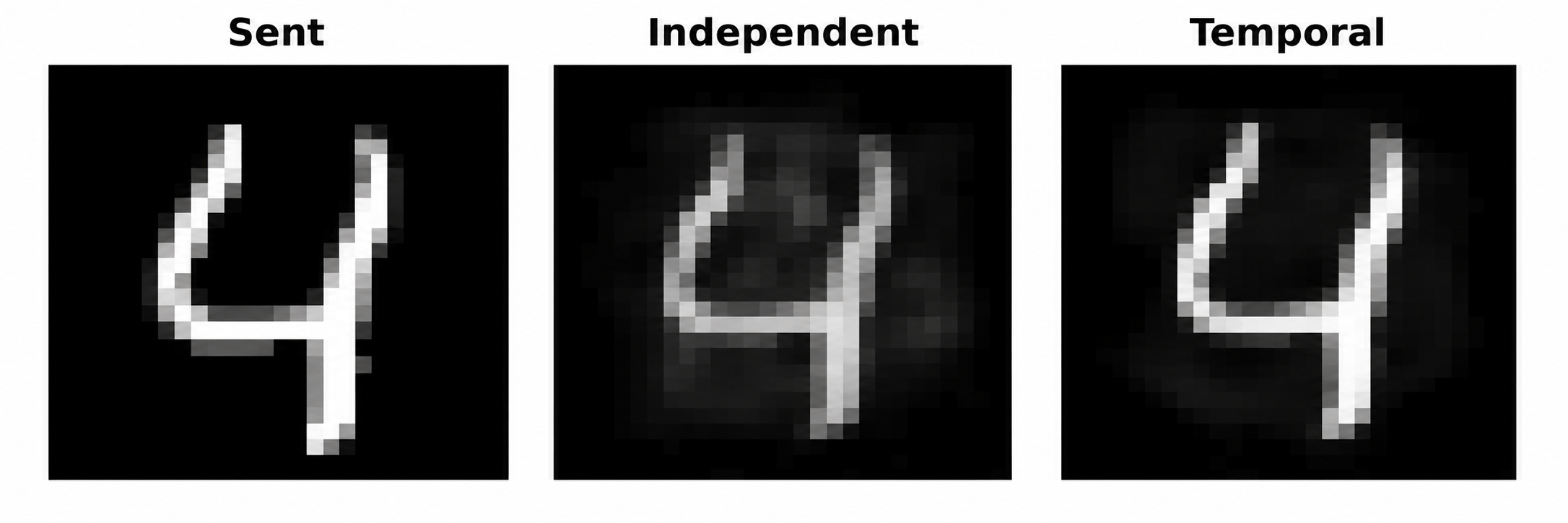}

    \includegraphics[
        width=0.95\linewidth,
        keepaspectratio
    ]{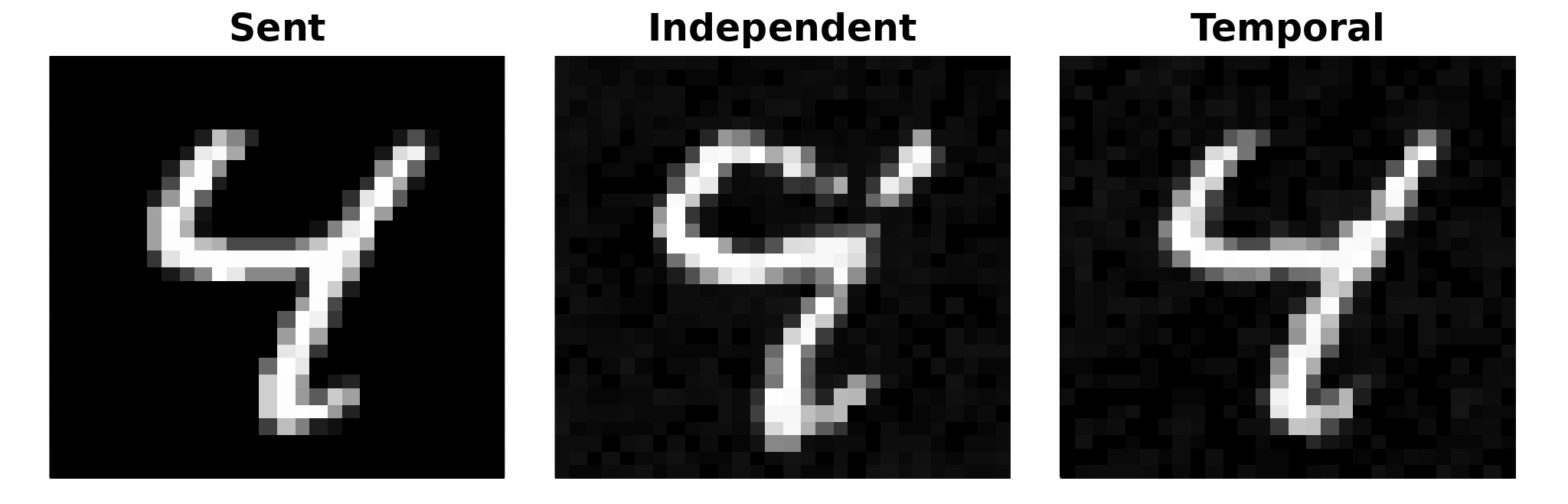}

    \caption{Sent images and the corresponding independent and temporal
    joint MIMO reconstructions.
    Top: Gaussian channel using the score-based model at SNR \(=10\) dB with \(\rH=0.4\).
    Bottom: \(\alpha\)-stable channel using LIM with
    \(\alpha=1.7\) at SNR \(=10\) dB with \(\rH=0.8\). These results correspond to the joint
    channel--image MIMO experiment in
    Section~\ref{subsec:experiments_joint_mimo}.}
    \label{fig:appendix_continuous_mimo_images}
    \vspace{-3mm}
\end{figure}

Each sequence contains \(K=10\) MNIST images. Figure~\ref{fig:mimo_snr_results} shows that temporal inference improves image MS-SSIM, channel NMSE (Gaussian) and \(\mathcal{E}_{\log,p}\) ($\alpha$-stable)  over independent inference as \(\rH\) increases. At \(\rH=0\), the methods coincide. Figure~\ref{fig:mimo_reconstructed_images} depicts representative reconstructed images for the two reconstruction methods and shows that temporal reconstruction becomes increasingly beneficial over independent one as \(\rH\) increases for both Gaussian and $\alpha$-stable channels. Similar trends are observed for the score-based and LIM models in Fig.~\ref{fig:appendix_continuous_mimo}, with representative continuous-time reconstructions shown in Fig.~\ref{fig:appendix_continuous_mimo_images}. 

\subsubsection*{Model mismatch}
\begin{figure}[!t]
\centering
\includegraphics[width=\linewidth]
{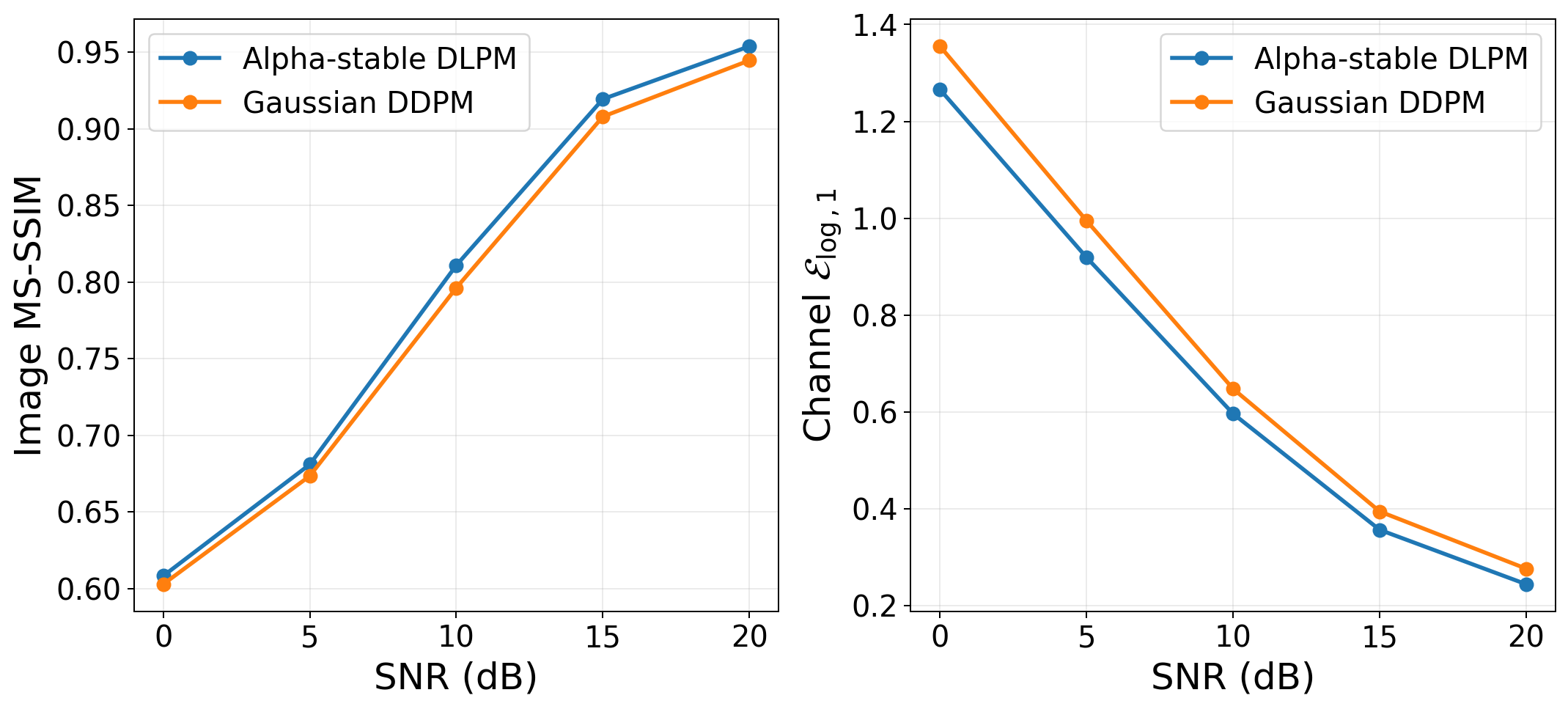}
\caption{Image MS-SSIM and channel loss vs. SNR for $\alpha$-stable joint channel--image MIMO reconstruction using matched DLPM and mismatched DDPM for \(\alpha=1.7\).}
\label{fig:stable_mimo_model_comparison}
\vspace{1mm}
\end{figure}

\begin{figure}[!t]
\centering
\includegraphics[
  width=\linewidth,
  keepaspectratio
]
{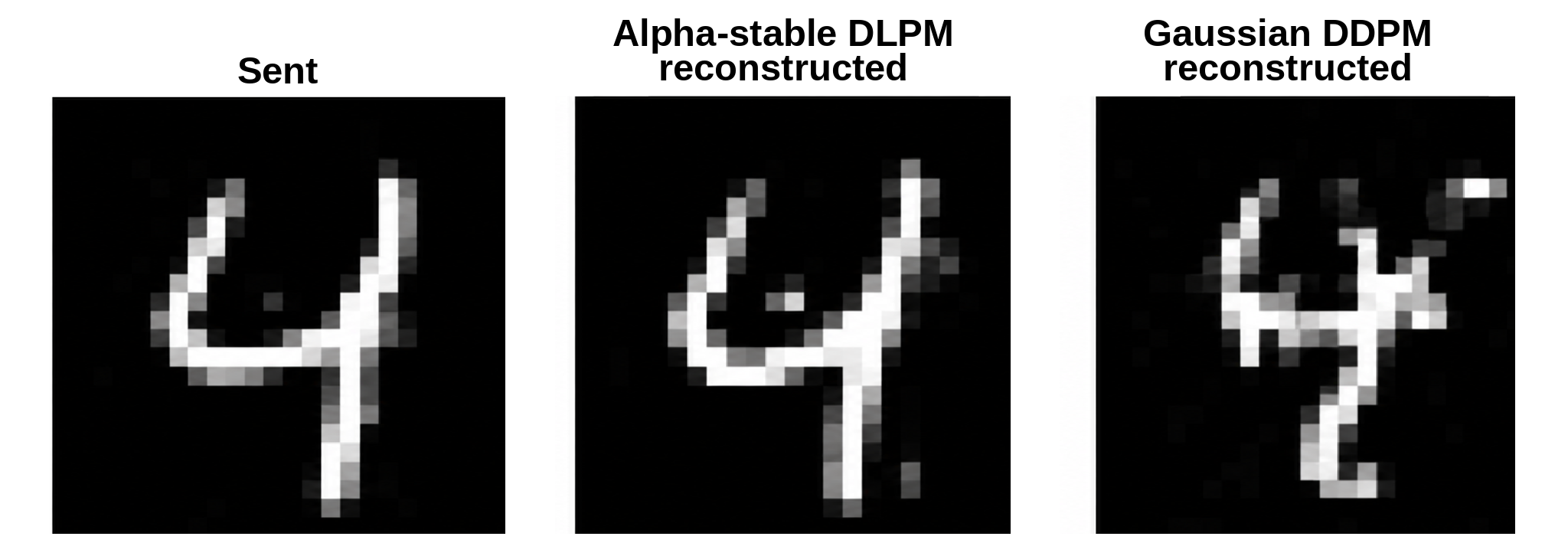}
\caption{Sent image on $\alpha$-stable channel with joint MIMO reconstructions obtained using matched DLPM and mismatched Gaussian DDPM for \(\alpha=1.7\) and SNR \(=15\) dB.}
\label{fig:stable_mimo_model_images}
\vspace{-3mm}
\end{figure}

\begin{figure}[!ht]
\centering
\includegraphics[width=\linewidth]
{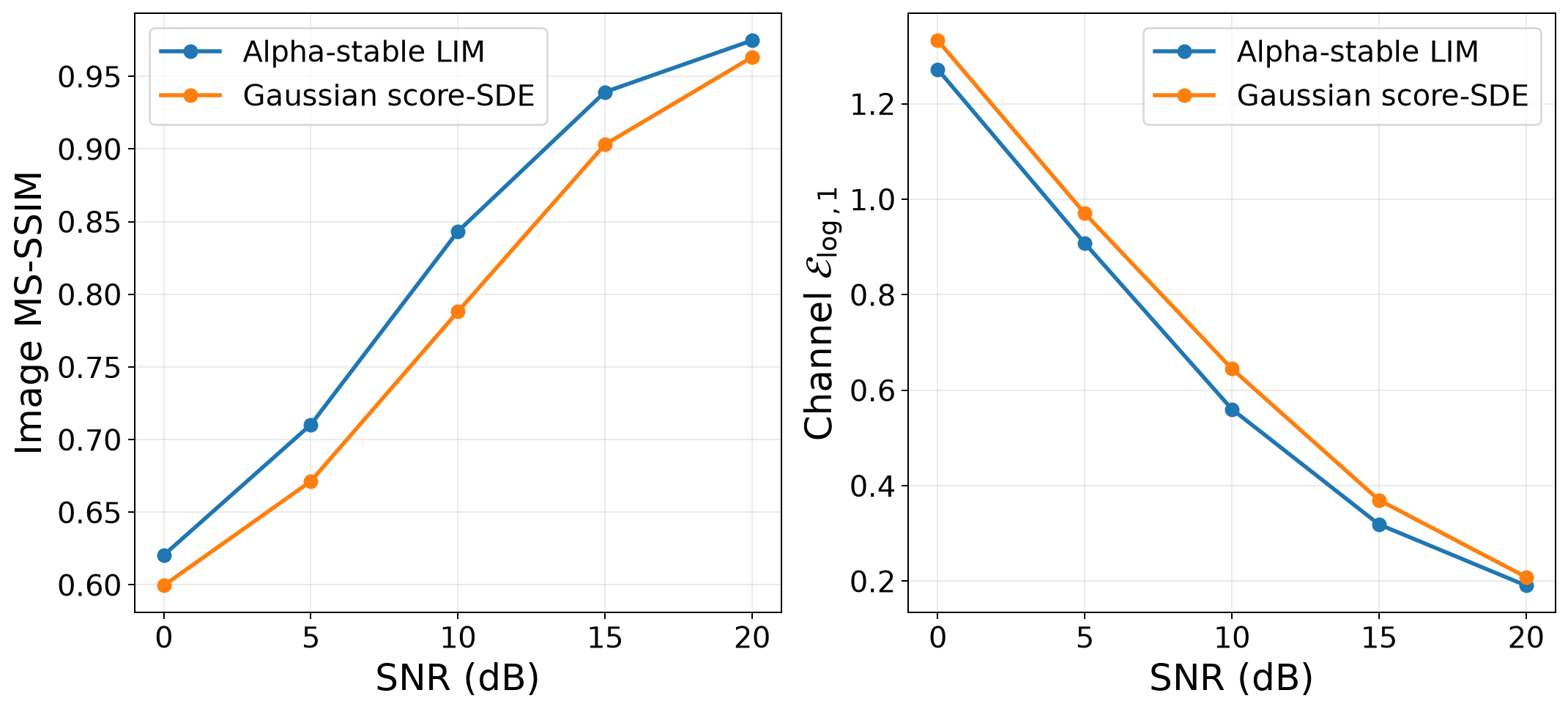}
\caption{Image MS-SSIM and channel loss vs. SNR for $\alpha$-stable joint
channel--image MIMO reconstruction using matched LIM and mismatched Gaussian
score-based model for \(\alpha=1.7\). This corresponds to the model-mismatch
experiment in Section~\ref{subsec:experiments_joint_mimo}.}
\label{fig:appendix_continuous_model_mismatch}
\end{figure}

\begin{figure}[ht!]
\centering
\includegraphics[
  width=\linewidth,
  keepaspectratio
]
{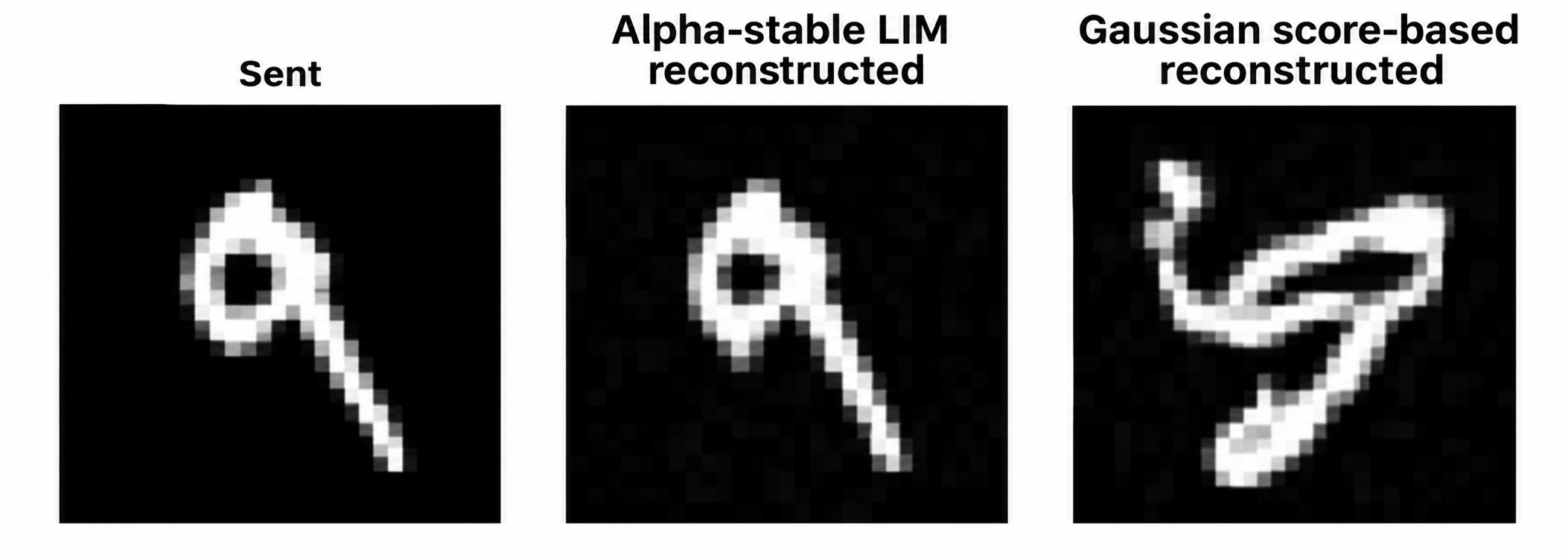}
\caption{Sent image on $\alpha$-stable channel with joint MIMO
reconstructions obtained using matched LIM and mismatched Gaussian
score-based model for \(\alpha=1.7\) and SNR \(=15\) dB. These results
correspond to the model-mismatch experiment in
Section~\ref{subsec:experiments_joint_mimo}.}
\label{fig:appendix_continuous_model_images}
\end{figure}

Matched $\alpha$-stable channel priors are compared with Gaussian channel priors trained and evaluated on the same $\alpha$-stable channel data with $\rho=0$. DLPM is compared with Gaussian DDPM in Figure~\ref{fig:stable_mimo_model_comparison}, while LIM is compared with the Gaussian score-based model. It is apparent that DLPM provides lower channel loss and higher image MS-SSIM than the Gaussian DDPM. Figure~\ref{fig:stable_mimo_model_images} shows a representative reconstructed image for the two channel priors. Similar trends are observed when comparing LIM with the Gaussian score-based model in Fig.~\ref{fig:appendix_continuous_model_mismatch}, with representative reconstructions shown in Fig.~\ref{fig:appendix_continuous_model_images}. 

\section{Discussion}
In this work, we leveraged time-dependence in channel statistics in diffusion-based channel estimation. Propagating previous estimates becomes more informative and improves both signal reconstruction and blind-MIMO recovery. This advantage is common to both Gaussian and $\alpha$-stable models.

The channel-model comparison separately shows the importance of matching the diffusion noise family with the underlying data distribution. For Gaussian and $\alpha$-stable channel models trained on the same
heavy-tailed data and using the same observations and likelihood, the improved channel loss and image quality obtained by DLPM, with similar behavior for LIM, therefore suggest that stable perturbations provide a better-matched prior for $\alpha$-stable channels. 

In the blind MIMO setting, 
while the assumption of perfect knowledge of the initial channel estimate is motivated by prior training, this might require a prohibitively long training period. Also, considering the previous channel estimate as exact when constructing the temporal prior is suboptimal. Future work could focus on reducing channel uncertainty propagation over time and the need for a long initial training stage.

\bibliographystyle{IEEEtran}
\bibliography{references}

@inproceedings{ho2020ddpm,
  author    = {Ho, Jonathan and Jain, Ajay and Abbeel, Pieter},
  title     = {Denoising Diffusion Probabilistic Models},
  booktitle = {Advances in Neural Information Processing Systems (NeurIPS)},
  year      = {2020}
}

@inproceedings{song2021scorebased,
  author    = {Song, Yang and Sohl-Dickstein, Jascha and Kingma, Diederik P. and Kumar, Abhishek and Ermon, Stefano and Poole, Ben},
  title     = {Score-Based Generative Modeling Through Stochastic Differential Equations},
  booktitle = {Proc. Int. Conf. Learn. Represent. (ICLR)},
  year      = {2021}
}

@inproceedings{chung2023dps,
  author    = {Chung, Hyungjin and Kim, Jeongsol and McCann, Michael T. and Klasky, Marc L. and Ye, Jong Chul},
  title     = {Diffusion Posterior Sampling for General Noisy Inverse Problems},
  booktitle = {Proc. Int. Conf. Learn. Represent. (ICLR)},
  year      = {2023}
}

@article{mohseni2026piecewise,
  author  = {Mohseni-Sehdeh, S. and Saad, Walid and Sakaguchi, Kei and Yu, T.},
  title   = {Diffusion Models for Solving Inverse Problems via Posterior Sampling With Piecewise Guidance},
  journal = {Trans. Mach. Learn. Res.},
  year    = {2026}
}

@article{arvinte2023mimo,
  author  = {Arvinte, M. and Tamir, Jonathan I.},
  title   = {{MIMO} Channel Estimation Using Score-Based Generative Models},
  journal = {IEEE Trans. Wireless Commun.},
  volume  = {22},
  number  = {6},
  pages   = {3698--3713},
  month   = jun,
  year    = {2023}
}

@article{fesl2024diffusion,
  author  = {Fesl, Benedikt and Baur, M. and Strasser, F. and Joham, Michael and Utschick, Wolfgang},
  title   = {Diffusion-Based Generative Prior for Low-Complexity {MIMO} Channel Estimation},
  journal = {IEEE Wireless Commun. Lett.},
  volume  = {13},
  number  = {12},
  pages   = {3493--3497},
  month   = dec,
  year    = {2024}
}

@article{zilberstein2023joint,
  author  = {Zilberstein, N. and Swami, Ananthram and Segarra, Santiago},
  title   = {Joint Channel Estimation and Data Detection in Massive {MIMO} Systems Based on Diffusion Models},
  journal = {arXiv preprint arXiv:2311.10311},
  year    = {2023}
}

@article{jiang2025blind,
  author  = {Jiang, H. and Yuan, X. and Huang, Y. and Guo, Q.},
  title   = {Blind {MIMO} Semantic Communication via Parallel Variational Diffusion: A Completely Pilot-Free Approach},
  journal = {arXiv preprint arXiv:2510.27043},
  year    = {2025}
}

@inproceedings{shariatian2025dlpm,
  author    = {Shariatian, D. and {\c{S}}im{\c{s}}ekli, Umut and Durmus, Alain O.},
  title     = {Heavy-Tailed Diffusion With Denoising L{\'e}vy Probabilistic Models},
  booktitle = {Proc. Int. Conf. Learn. Represent. (ICLR)},
  year      = {2025}
}

@book{samorodnitsky1994stable,
  author    = {Samorodnitsky, Gennady and Taqqu, Murad S.},
  title     = {Stable Non-Gaussian Random Processes},
  publisher = {Chapman \& Hall},
  year      = {1994}
}

@book{kolmo1968,
  author    = {Gnedenko, B. V. and Kolmogorov, A. N.},
  title     = {Limit Distributions for Sums of Independent Random Variables},
  address   = {Reading, MA, USA},
  publisher = {Addison-Wesley Publishing Company},
  year      = {1968}
}

@article{fahs2026,
  author  = {Fahs, Jihad},
  title   = {On Performance Analysis in Heavy-Tailed (In)Dependent Wireless Channels},
  journal = {IEEE Commun. Lett.},
  volume  = {30},
  pages   = {2278--2282},
  year    = {2026},
  doi     = {10.1109/LCOMM.2026.3704058}
}

@inproceedings{wang2003msssim,
  author    = {Wang, Zhou and Simoncelli, Eero P. and Bovik, Alan C.},
  title     = {Multiscale Structural Similarity for Image Quality Assessment},
  booktitle = {Proc. Asilomar Conf. Signals, Syst. Comput.},
  year      = {2003}
}

@inproceedings{yoon2023lim,
  author    = {Yoon, E. B. and Park, K. and Kim, S. and Lim, S.},
  title     = {Score-Based Generative Models With L{\'e}vy Processes},
  booktitle = {Advances in Neural Information Processing Systems},
  volume    = {36},
  year      = {2023}
}

@article{lecun1998mnist,
  author  = {LeCun, Yann and Bottou, L{\'e}on and Bengio, Yoshua and Haffner, Patrick},
  title   = {Gradient-Based Learning Applied to Document Recognition},
  journal = {Proc. IEEE},
  volume  = {86},
  number  = {11},
  pages   = {2278--2324},
  year    = {1998}
}

@article{mcculloch1986simple,
  author  = {McCulloch, J. Huston},
  title   = {Simple Consistent Estimators of Stable Distribution Parameters},
  journal = {Commun. Stat. Simul. Comput.},
  volume  = {15},
  number  = {4},
  pages   = {1109--1136},
  year    = {1986},
  doi     = {10.1080/03610918608812563}
}

@inproceedings{fahs2026linearity,
  author    = {Fahs, Jihad and Abou-Faycal, Ibrahim},
  title     = {On the Linearity of Conditional Mean Estimators of Infinite-Variance Variables},
  booktitle = {Proc. IEEE Int. Symp. Inf. Theory (ISIT)},
  pages     = {1--6},
  year      = {2026},
  doi       = {10.1109/ISIT62367.2026.11653965}
}

@book{doucet2001sequential,
  editor    = {Doucet, Arnaud and de Freitas, Nando and Gordon, Neil},
  title     = {Sequential Monte Carlo Methods in Practice},
  address   = {New York, NY, USA},
  publisher = {Springer},
  year      = {2001}
}

@inproceedings{aub-htp,
  author    = {Tabbara, Wael and Sharafeddine, Sharafeddine and El-Hajj, Ahmad and Fahs, Jihad and Abou-Faycal, Ibrahim},
  title     = {Density Functions and Random Number Generators of {$\alpha$}-Stable Distributions},
  booktitle = {Proc. Python in Science Conf. (SciPy)},
  year      = {2026},
  url       = {https://proceedings.scipy.org/articles/eykk7863}
}

\appendices

\section{DDPM}
\label{app:ddpm-comprehensive}

\subsection{Forward and Reverse Processes}

A DDPM defines the forward process
\begin{equation*}
q(\vX_t\mid\vX_{t-1}=\vx_{t-1})
=
\mathcal{N}\left(
\sqrt{\alpha_t}\vx_{t-1},
\beta_t\mat{I}_d
\right),
\quad \alpha_t=1-\beta_t,
\end{equation*}
whose marginal, with
$\bar{\alpha}_t=\prod_{s=1}^t\alpha_s$, is
\begin{equation}
\vX_t=
\sqrt{\bar{\alpha}_t}\vX_0+
\sqrt{1-\bar{\alpha}_t}\vepsilon_t,
\quad
\vepsilon_t\sim\mathcal{N}(\vzeros,\mat{I}_d).
\label{eq:ddpm_forward_marginal}
\end{equation}
The exact reverse posterior conditioned on $\vX_0=\vx_0$ is
\begin{equation*}
q(\vX_{t-1}\mid\vX_t=\vx_t,\vX_0=\vx_0)
=
\mathcal{N}(\widetilde{\vmu}_t,
\widetilde{\beta}_t\mat{I}_d),
\end{equation*}
where
\begin{align*}
\widetilde{\vmu}_t
&=
\frac{\sqrt{\bar{\alpha}_{t-1}}\beta_t}
{1-\bar{\alpha}_t}\vx_0
+
\frac{\sqrt{\alpha_t}(1-\bar{\alpha}_{t-1})}
{1-\bar{\alpha}_t}\vx_t,\\
\widetilde{\beta}_t
&=
\frac{1-\bar{\alpha}_{t-1}}
{1-\bar{\alpha}_t}\beta_t.
\end{align*}
The learned reverse transition replaces the unknown
perturbation by $\vepsilon_\theta(\vx_t,t)$ such that 
\begin{equation*}
p_\theta(\vX_{t-1}\mid\vX_t=\vx_t)
=
\mathcal{N}\left(
\vmu_\theta(\vx_t,t),
\widetilde{\beta}_t\mat{I}_d
\right),
\end{equation*}
with
\begin{equation}
\vmu_\theta(\vx_t,t)
=
\frac{1}{\sqrt{\alpha_t}}
\left(
\vx_t-
\frac{\beta_t}{\sqrt{1-\bar{\alpha}_t}}
\vepsilon_\theta(\vx_t,t)
\right).
\label{eq:ddpm_reverse_mean_parameterization}
\end{equation}
The corresponding clean estimate is
\begin{equation*}
\widehat{\vx}_0(\vx_t,t)
=
\frac{\vx_t-\sqrt{1-\bar{\alpha}_t}
\vepsilon_\theta(\vx_t,t)}
{\sqrt{\bar{\alpha}_t}}.
\end{equation*}
Starting from the terminal marginal, reverse sampling
uses
$\vx_{t-1}=\vmu_\theta(\vx_t,t)
+\sqrt{\widetilde{\beta}_t}\vZ_t$,
where $\vZ_t\sim\mathcal{N}(\vzeros,\mat{I}_d)$
and $\vZ_1=\vzeros$.

\subsection{Canonical Gaussian Adaptation}
\label{subsec:appendix_ddpm_canonical_gaussian_adaptation}

We train a single DDPM on the canonical distribution
$\vZ_0\sim\mathcal{N}(\vzeros,\mat{I}_d)$ and adapt its
noise predictor to $\vX_0\sim\mathcal{N}(\vmu,\mat{\Sigma})$
without retraining. From \eqref{eq:ddpm_forward_marginal},
\[
\vX_t\sim\mathcal{N}
\left(\sqrt{\bar{\alpha}_t}\,\vmu,\mat{C}_t\right),
\qquad
\mat{C}_t=\bar{\alpha}_t\mat{\Sigma}
+(1-\bar{\alpha}_t)\mat{I}_d.
\]
Let $\mat{C}_t=\mat{R}_t\mat{R}_t^{\mathsf T}$
be its Cholesky factorization. The transformed input
\[
\vU_t=\mat{R}_t^{-1}
\left(\vX_t-\sqrt{\bar{\alpha}_t}\,\vmu\right)
\]
satisfies $\vU_t\sim\mathcal{N}(\vzeros,\mat{I}_d)$.

\begin{lemma}
Let $\vepsilon_\theta$ be the MSE-optimal noise
predictor trained on the canonical distribution
$\mathcal{N}(\vzeros,\mat{I}_d)$. Then, the
MSE-optimal noise predictor for
$\vX_0\sim\mathcal{N}(\vmu,\mat{\Sigma})$
is obtained without retraining as
\begin{align*}
\widehat{\vepsilon}_{\vmu,\mat{\Sigma}}(\vX_t,t)
=
\mat{R}_t^{-\mathsf T}\vepsilon_\theta(\vU_t,t).
\end{align*}
\end{lemma}

\begin{proof}
Under the squared-error objective, the optimal
predictor is the conditional expectation
$\mathbb{E}[\vepsilon_t\mid\vX_t]$.
Since $\vX_0$ and $\vepsilon_t$ are independent
Gaussian vectors,
\[
\begin{bmatrix}\vX_0\\\vepsilon_t\end{bmatrix}
\sim\mathcal{N}\left(
\begin{bmatrix}\vmu\\\vzeros\end{bmatrix},
\begin{bmatrix}
\mat{\Sigma}&\mat{0}\\
\mat{0}&\mat{I}_d
\end{bmatrix}
\right).
\]
By \eqref{eq:ddpm_forward_marginal},
$(\vepsilon_t,\vX_t)$ is a linear transformation
of $(\vX_0,\vepsilon_t)$ and is therefore
jointly Gaussian. Its cross-covariance is
\begin{align*}
\operatorname{Cov}(\vepsilon_t,\vX_t)
&=\sqrt{\bar{\alpha}_t}
\operatorname{Cov}(\vepsilon_t,\vX_0)\\
&\quad+\sqrt{1-\bar{\alpha}_t}
\operatorname{Cov}(\vepsilon_t)\\
&=\sqrt{1-\bar{\alpha}_t}\,\mat{I}_d,
\end{align*}
where independence makes the first term zero.

For jointly Gaussian vectors, the conditional
expectation is
\[
\begin{aligned}
\mathbb{E}[\vepsilon_t\mid\vX_t]
&=\mathbb{E}[\vepsilon_t]
+\operatorname{Cov}(\vepsilon_t,\vX_t)\\
&\quad\cdot\operatorname{Cov}(\vX_t)^{-1}
\left(\vX_t-\mathbb{E}[\vX_t]\right).
\end{aligned}
\]
Using $\mathbb{E}[\vepsilon_t]=\vzeros$,
$\mathbb{E}[\vX_t]=\sqrt{\bar{\alpha}_t}\,\vmu$,
$\operatorname{Cov}(\vepsilon_t,\vX_t)
=\sqrt{1-\bar{\alpha}_t}\,\mat{I}_d$, and
$\operatorname{Cov}(\vX_t)=\mat{C}_t$, we obtain
\begin{align*}
\mathbb{E}[\vepsilon_t\mid\vX_t]
&=\sqrt{1-\bar{\alpha}_t}\,
\mat{C}_t^{-1}
\left(\vX_t-\sqrt{\bar{\alpha}_t}\,\vmu\right)\\
&=\sqrt{1-\bar{\alpha}_t}\,
\mat{R}_t^{-\mathsf T}\mat{R}_t^{-1}
\left(\vX_t-\sqrt{\bar{\alpha}_t}\,\vmu\right)\\
&=\sqrt{1-\bar{\alpha}_t}\,
\mat{R}_t^{-\mathsf T}\vU_t.
\end{align*}

For the canonical model,
$\vZ_t\sim\mathcal{N}(\vzeros,\mat{I}_d)$ and
$\operatorname{Cov}(\vepsilon_t,\vZ_t)
=\sqrt{1-\bar{\alpha}_t}\mat{I}_d$.
Hence, its MSE-optimal predictor satisfies
\[
\vepsilon_\theta(\vZ_t,t)
=\mathbb{E}[\vepsilon_t\mid\vZ_t]
=\sqrt{1-\bar{\alpha}_t}\,\vZ_t.
\]
Evaluating it at $\vZ_t=\vU_t$ gives
\[
\begin{aligned}
\mat{R}_t^{-\mathsf T}\vepsilon_\theta(\vU_t,t)
&=\sqrt{1-\bar{\alpha}_t}\,
\mat{R}_t^{-\mathsf T}\vU_t\\
&=\mathbb{E}[\vepsilon_t\mid\vX_t],
\end{aligned}
\]
which establishes
\eqref{eq:ddpm_gaussian_adaptation}.
\end{proof}

\subsection{Conditional Sampling Reconstruction}

We consider the linear observation model in \eqref{eq:linear_gaussian_model}, with the posterior mean in \eqref{eq:linear_gaussian_mean} and posterior covariance in \eqref{eq:linear_gaussian_covariance}.
Conditional DDPM sampling incorporates the observation before the reverse
process: the Gaussian posterior becomes the target Gaussian law, and the
canonical adaptation in \eqref{eq:ddpm_gaussian_adaptation} is applied using
(\eqref{eq:linear_gaussian_mean},\eqref{eq:linear_gaussian_covariance}).
Consequently, no likelihood gradient is added during the reverse steps.
We generate $P$ samples from the conditional law and select the one
that maximizes the likelihood of Gaussian observation.

In this work, conditional sampling is used only for the Gaussian
channel-only experiment, with $\vxi_0=\vH_k$ and
$\mat{A}=\mat{I}_d$.
For the independent conditional method, $(\vmu,\mat{\Sigma})$
in \eqref{eq:linear_gaussian_mean} and
\eqref{eq:linear_gaussian_covariance} is the stationary channel prior
in \eqref{eq:gaussian_channel_dynamics}; for the temporal conditional
method, it is the predictive prior in
\eqref{eq:gaussian_predictive_channel_prior}.

Algorithm~\ref{alg:ddpm_conditional} summarizes this conditional
reconstruction procedure.

\begin{algorithm}[t]
\caption{DDPM Conditional Sampling Reconstruction}
\label{alg:ddpm_conditional}
\begin{algorithmic}[1]
\REQUIRE $\vy$, $\mat{A}$, $(\vmu,\mat{\Sigma})$, $\sigma_n^2$, $\vepsilon_\theta$, DDPM schedule, $P$
\STATE Compute $(\vmu_{\mathrm{post}},\mat{\Sigma}_{\mathrm{post}})$ using
\eqref{eq:linear_gaussian_mean} and \eqref{eq:linear_gaussian_covariance}
\FOR{$p=1,\ldots,P$}
    \STATE Draw $\vxi_T^{(p)}$ from the terminal marginal of the posterior forward process
    \FOR{$t=T,\ldots,1$}
        \STATE $\mat{C}_{\mathrm{post},t}\gets
        \bar{\alpha}_t\mat{\Sigma}_{\mathrm{post}}+(1-\bar{\alpha}_t)\mat{I}_d$
        \STATE $\mat{R}_t\mat{R}_t^{\mathsf T}\gets\mat{C}_{\mathrm{post},t}$
        \STATE $\vU_t^{(p)}\gets\mat{R}_t^{-1}
        (\vxi_t^{(p)}-\sqrt{\bar{\alpha}_t}\vmu_{\mathrm{post}})$
        \STATE $\widehat{\vepsilon}_t^{(p)}\gets
        \mat{R}_t^{-\mathsf T}\vepsilon_\theta(\vU_t^{(p)},t)$
        \STATE Compute $\vmu_{\theta,t}^{(p)}$ using
        \eqref{eq:ddpm_reverse_mean_parameterization}
        with $\vX_t=\vxi_t^{(p)}$ and
        $\vepsilon_\theta=\widehat{\vepsilon}_t^{(p)}$
        \STATE Advance $\vxi_t^{(p)}$ to $\vxi_{t-1}^{(p)}\gets
        \vmu_{\theta,t}^{(p)}+\sqrt{\widetilde{\beta}_t}\vZ_t^{(p)}$
        \COMMENT{$\vZ_1^{(p)}=\vzeros$}
    \ENDFOR
\ENDFOR
\STATE $p^\star\gets\arg\max_{1\leq p\leq P}p(\vy\mid\vxi_0^{(p)})$
\RETURN $\widehat{\vxi}_0=\vxi_0^{(p^\star)}$
\end{algorithmic}
\end{algorithm}

\subsection{Likelihood-Guided Reconstruction}

We consider the general observation model
\eqref{eq:generic_observation_model}, where each
unknown $\vxi_0^{(j)}$ has a separately trained
DDPM prior. This formulation includes both
channel-only estimation in
\eqref{eq:channel_only_model} and joint MIMO
reconstruction in \eqref{eq:joint_mimo_problem}.

For Gaussian channel-only estimation, we set $J=1$, $\vxi_0^{(1)}=\vH_k$, and $f(\vxi_0^{(1)})=\vxi_0^{(1)}$.
The independent method uses the stationary Gaussian
channel prior in \eqref{eq:gaussian_channel_dynamics},
whereas the temporal method uses the Gaussian predictive prior
in \eqref{eq:gaussian_predictive_channel_prior}.

For blind joint channel--image MIMO reconstruction, we set $J=2$, $\vxi_0^{(1)}=\operatorname{vec}(\mat{H}_k)$, $\vxi_0^{(2)}=\operatorname{vec}(\mat{D}_k)$, and $f(\vxi_0^{(1)},\vxi_0^{(2)})=\operatorname{vec}(\mat{H}_k f_\gamma(\mat{D}_k))$.
The channel and image have separately trained
DDPM priors, coupled through the MIMO likelihood.
The independent method uses the stationary Gaussian
channel prior in \eqref{eq:gaussian_channel_dynamics},
whereas the temporal method uses the predictive prior
in \eqref{eq:gaussian_mimo_temporal_prior}, treating
the previous channel estimate as exact.
The image prior is the same for both methods.

For each unknown and at every time step $t=T,\ldots,1$, the predicted clean state is
\begin{equation}
\widehat{\vxi}_{0,t}^{(j)}
=
\frac{\vxi_t^{(j)}
-\sqrt{1-\bar{\alpha}_t}
\widehat{\vepsilon}_t^{(j)}}
{\sqrt{\bar{\alpha}_t}},
\label{eq:appendix_general_clean_estimate}
\end{equation}
where $\widehat{\vepsilon}_t^{(j)}$ denotes
the adapted noise prediction.
The observation likelihood gradients are then
evaluated as
\begin{equation}
\vg_t^{(j)}
=
\nabla_{\vxi_t^{(j)}}
\log p\left(
\vy\mid
\widehat{\vxi}_{0,t}^{(1)},\ldots,
\widehat{\vxi}_{0,t}^{(J)}
\right).
\label{eq:appendix_general_guidance_gradient}
\end{equation}
The corresponding guided reverse update,
performed at every time step, is
\begin{equation}
\vxi_{t-1}^{(j)}
=
\vmu_{\theta,t}^{(j)}
+\eta_t^{(j)}\vg_t^{(j)}
+\sqrt{\widetilde{\beta}_t}\vZ_t^{(j)},
\label{eq:appendix_guided_update}
\end{equation}
where $\vmu_{\theta,t}^{(j)}$ is obtained from
\eqref{eq:ddpm_reverse_mean_parameterization},
$\eta_t^{(j)}$ is the guidance strength, and
$\vZ_1^{(j)}=\vzeros$.

For channel-only estimation, the likelihood
guides the channel reverse state. In blind MIMO
reconstruction, the channel and image reverse
states are both updated at every diffusion step
using gradients of the same joint observation
likelihood.

Among $P$ final candidates, we select the one
maximizing the observation likelihood.

Algorithm~\ref{alg:ddpm_guided} summarizes this likelihood-guided
reconstruction procedure.

\begin{algorithm}[t]
\caption{DDPM Likelihood-Guided Reconstruction}
\label{alg:ddpm_guided}
\begin{algorithmic}[1]
\REQUIRE $\vy$, Gaussian likelihood $p(\vy\mid\cdot)$,
$J$ DDPM priors, $\{\eta_t^{(j)}\}$, DDPM schedule, $P$
\FOR{$p=1,\ldots,P$}
    \STATE Draw $\{\vxi_T^{(j,p)}\}_{j=1}^{J}$ from their terminal marginals
    \FOR{$t=T,\ldots,1$}
        \FOR{$j=1,\ldots,J$}
            \STATE Compute adapted noise prediction $\widehat{\vepsilon}_t^{(j,p)}$
            \STATE Compute clean estimate $\widehat{\vxi}_{0,t}^{(j,p)}$
            using \eqref{eq:appendix_general_clean_estimate}
        \ENDFOR
        \FOR{$j=1,\ldots,J$}
            \STATE Compute likelihood gradient $\vg_t^{(j,p)}$
            using \eqref{eq:appendix_general_guidance_gradient}
            \STATE Advance $\vxi_t^{(j,p)}$ to $\vxi_{t-1}^{(j,p)}$
            using \eqref{eq:appendix_guided_update}
        \ENDFOR
    \ENDFOR
\ENDFOR
\STATE $p^\star\gets\arg\max_{1\leq p\leq P}
 p(\vy\mid\vxi_0^{(1,p)},\ldots,\vxi_0^{(J,p)})$
\RETURN $\{\vxi_0^{(j,p^\star)}\}_{j=1}^{J}$
\end{algorithmic}
\end{algorithm}

\section{Score-Based Diffusion}
\label{app:score-comprehensive}

\subsection{Forward and Reverse Processes}

A continuous score-based model defines the forward SDE
\begin{align*}
d\vX(t)
&=
\vect{b}(\vX(t),t)\,dt
+g(t)\,d\vW(t),
\end{align*}
where $\vW(t)$ is a standard Wiener process. We consider the linear
Gaussian perturbation kernel associated with this SDE,
\begin{equation}
\vX(t)
=
m(t)\vX_0+\sigma(t)\vepsilon(t),
\qquad
\vepsilon(t)\sim\mathcal{N}(\vzeros,\mat{I}_d).
\label{eq:score_forward_marginal_appendix}
\end{equation}
Let $p_t$ denote the marginal density of $\vX(t)$. The corresponding
reverse-time SDE is
\begin{multline}
d\vX(t)
=
\left[
\vect{b}(\vX(t),t)
-g^2(t)\nabla_{\vX(t)}\log p_t(\vX(t))
\right]dt \\
+g(t)\,d\overline{\vW}(t),
\label{eq:score_reverse_sde}
\end{multline}
which is solved from $t=T$ to $t=0$. The network
$\vs_\theta(\vX(t),t)$ approximates the score
$\nabla_{\vX(t)}\log p_t(\vX(t))$ and is substituted into
\eqref{eq:score_reverse_sde} during generation.

For the perturbation kernel in
\eqref{eq:score_forward_marginal_appendix}, the
clean estimate is
\begin{align*}
\widehat{\vX}_0(\vX(t),t)
&=
\frac{
\vX(t)+\sigma^2(t)\vs_\theta(\vX(t),t)
}{m(t)}.
\end{align*}
Starting from the terminal marginal, the reverse SDE is evaluated on a
decreasing numerical grid of $N_{\mathrm{grid}}$ intervals,
$T=t_{N_{\mathrm{grid}}}>\cdots>t_0=0$.

\subsection{Canonical Gaussian Adaptation}
\label{subsec:appendix_score_canonical_gaussian_adaptation}

We train a single score-based model on the canonical distribution
$\vZ_0\sim\mathcal{N}(\vzeros,\mat{I}_d)$ and adapt its score to
$\vX_0\sim\mathcal{N}(\vmu,\mat{\Sigma})$ without retraining.
From \eqref{eq:score_forward_marginal_appendix},
\[
\vX(t)
\sim
\mathcal{N}\!\left(
m(t)\vmu,\mat{C}(t)
\right),
\qquad
\mat{C}(t)
=
m^2(t)\mat{\Sigma}
+\sigma^2(t)\mat{I}_d.
\]
Let
$\mat{C}(t)=\mat{R}(t)\mat{R}^{\mathsf T}(t)$
be its Cholesky factorization. The transformed state
\[
\vU(t)
=
\mat{R}^{-1}(t)
\left(
\vX(t)-m(t)\vmu
\right)
\]
has the same canonical marginal as $\vZ(t)$.

\begin{lemma}
Let $\vs_\theta$ be the score trained on the canonical state $\vZ(t)$.
Then, the score for
$\vX_0\sim\mathcal{N}(\vmu,\mat{\Sigma})$
is obtained without retraining as
\begin{align*}
\widehat{\vs}_{\vmu,\mat{\Sigma}}(\vX(t),t)
=
\mat{R}^{-\mathsf T}(t)
\vs_\theta(\vU(t),t).
\end{align*}
\end{lemma}

\begin{proof}
Let $p_{\vX(t)}(\vx(t))$ and $p_{\vU(t)}(\vu(t))$ denote the densities of
$\vX(t)$ and $\vU(t)$. Since
\(
\vU(t)
=
\mat{R}^{-1}(t)
\left(
\vX(t)-m(t)\vmu
\right)
\) and \(
\vu(t)
=
\mat{R}^{-1}(t)
\left(
\vx(t)-m(t)\vmu
\right),
\) 
the change-of-variables formula gives
\[
p_{\vX(t)}(\vx(t))
=
\frac{1}{|\det\mat{R}(t)|}
p_{\vU(t)}\!\left(\vu(t)\right).
\]
Hence,
\[
\log p_{\vX(t)}(\vx(t))
=
-\log|\det\mat{R}(t)|
+\log p_{\vU(t)}(\vu(t)).
\]
Because \(
\nabla_{\vx(t)}
\log|\det\mat{R}(t)|
=
\vzeros
\), applying the chain rule therefore gives
\begin{align*}
\nabla_{\vx(t)}
\log p_{\vX(t)}(\vx(t))
&=
\mat{R}^{-\mathsf T}(t)
\nabla_{\vu(t)}
\log p_{\vU(t)}(\vu(t)).
\end{align*}
For the canonical model
$\vZ(t)\sim\mathcal N(\vzeros,\mat I_d)$,
the score function is
\(
\vs_\theta(\vZ(t),t)
\approx
\left.
\nabla_{\vu(t)}\log p_{\vU(t)}(\vu(t))
\right|_{\vu(t)=\vZ(t)}.
\)
Evaluating it at $\vZ(t)=\vU(t)$ gives
\[
\begin{aligned}
\vs_\theta(\vU(t),t)
&=
\left.
\nabla_{\vu(t)}\log p_{\vU(t)}(\vu(t))
\right|_{\vu(t)=\vU(t)}
\\[6pt]
&=
\mat{R}^{\mathsf T}(t)
\left.
\nabla_{\vx(t)}\log p_{\vX(t)}(\vx(t))
\right|_{\vx(t)=\vX(t)}
\\[6pt]
&=
\mat{R}^{\mathsf T}(t)
\vs_\theta(\vX(t),t).
\end{aligned}
\]

which establishes
\eqref{eq:score_gaussian_adaptation}.
\end{proof}

\subsection{Conditional Sampling Reconstruction}

We consider the linear observation model in \eqref{eq:linear_gaussian_model}, with the posterior mean in \eqref{eq:linear_gaussian_mean} and posterior covariance in \eqref{eq:linear_gaussian_covariance}.
Conditional score-based sampling incorporates the observation before the reverse
process: the Gaussian posterior becomes the target Gaussian law, and the
canonical adaptation in \eqref{eq:score_gaussian_adaptation} is applied using
(\eqref{eq:linear_gaussian_mean},\eqref{eq:linear_gaussian_covariance}).
Consequently, no likelihood gradient is added during the reverse steps.
We generate $P$ samples from the conditional law and select the one
maximizing the Gaussian observation likelihood.

In this work, conditional sampling is used only for the Gaussian
channel-only experiment, with $\vxi_0=\vH_k$ and
$\mat{A}=\mat{I}_d$.
For the independent conditional method, $(\vmu,\mat{\Sigma})$
in \eqref{eq:linear_gaussian_mean} and
\eqref{eq:linear_gaussian_covariance} is the stationary channel prior
in \eqref{eq:gaussian_channel_dynamics}; for the temporal conditional
method, it is the predictive prior in
\eqref{eq:gaussian_predictive_channel_prior}.

Algorithm~\ref{alg:score_conditional} summarizes this conditional
reconstruction procedure.

\begin{algorithm}[t]
\caption{Score-Based Conditional Sampling Reconstruction}
\label{alg:score_conditional}
\begin{algorithmic}[1]
\REQUIRE $\vy$, $\mat{A}$, $(\vmu,\mat{\Sigma})$, $\sigma_n^2$, $\vs_\theta$, SDE coefficients, reverse grid, $P$
\STATE Compute $(\vmu_{\mathrm{post}},\mat{\Sigma}_{\mathrm{post}})$ using
\eqref{eq:linear_gaussian_mean} and \eqref{eq:linear_gaussian_covariance}
\FOR{$p=1,\ldots,P$}
    \STATE Draw $\vxi^{(p)}(T)$ from the terminal marginal of the posterior forward process
    \FOR{$\ell=N_{\mathrm{grid}},\ldots,1$}
        \STATE $t\gets t_\ell$
        \STATE $\mat{C}_{\mathrm{post}}(t)\gets
        m^2(t)\mat{\Sigma}_{\mathrm{post}}+\sigma^2(t)\mat{I}_d$
        \STATE $\mat{R}(t)\mat{R}^{\mathsf T}(t)\gets\mat{C}_{\mathrm{post}}(t)$
        \STATE $\vU^{(p)}(t)\gets\mat{R}^{-1}(t)
        (\vxi^{(p)}(t)-m(t)\vmu_{\mathrm{post}})$
        \STATE $\widehat{\vs}^{(p)}(t)\gets
        \mat{R}^{-\mathsf T}(t)\vs_\theta(\vU^{(p)}(t),t)$
        \STATE Advance $\vxi^{(p)}(t)$ to $\vxi^{(p)}(t_{\ell-1})$
        using \eqref{eq:score_reverse_sde} with score $\widehat{\vs}^{(p)}(t)$
    \ENDFOR
\ENDFOR
\STATE $p^\star\gets\arg\max_{1\leq p\leq P}p(\vy\mid\vxi^{(p)}(0))$
\RETURN $\widehat{\vxi}_0=\vxi^{(p^\star)}(0)$
\end{algorithmic}
\end{algorithm}

\subsection{Likelihood-Guided Reconstruction}

We consider the general observation model
\eqref{eq:generic_observation_model}, where each
unknown $\vxi_0^{(j)}$ has a separately trained
score-based diffusion prior. This formulation includes both
channel-only estimation in
\eqref{eq:channel_only_model} and joint MIMO
reconstruction in \eqref{eq:joint_mimo_problem}.

For Gaussian channel-only estimation, we set
$J=1$, $\vxi_0^{(1)}=\vH_k$, and
$f(\vxi_0^{(1)})=\vxi_0^{(1)}$.
The independent method uses the stationary Gaussian
channel prior in \eqref{eq:gaussian_channel_dynamics},
whereas the temporal method uses the Gaussian predictive prior
in \eqref{eq:gaussian_predictive_channel_prior}.

For blind joint channel--image MIMO reconstruction, we set
$J=2$,
$\vxi_0^{(1)}=\operatorname{vec}(\mat{H}_k)$,
$\vxi_0^{(2)}=\operatorname{vec}(\mat{D}_k)$, and
$f(\vxi_0^{(1)},\vxi_0^{(2)})
=\operatorname{vec}(\mat{H}_k f_\gamma(\mat{D}_k))$.
The channel and image have separately trained
score-based diffusion priors, coupled through the MIMO likelihood.
The independent method uses the stationary Gaussian
channel prior in \eqref{eq:gaussian_channel_dynamics},
whereas the temporal method uses the predictive prior
in \eqref{eq:gaussian_mimo_temporal_prior}, treating
the previous channel estimate as exact.
The image prior is the same for both methods.

For each unknown and at every reverse time $t$,
the predicted clean state is
\begin{equation}
\widehat{\vxi}_0^{(j)}(t)
=
\frac{
\vxi^{(j)}(t)
+\sigma_j^2(t)
\widehat{\vs}^{(j)}(\vxi^{(j)}(t),t)
}{m_j(t)},
\label{eq:score_general_clean_estimate}
\end{equation}
where $\widehat{\vs}^{(j)}$ denotes
the adapted score estimate.
The observation likelihood gradients are then
evaluated as
\begin{equation}
\vg^{(j)}(t)
=
\nabla_{\vxi^{(j)}(t)}
\log p\left(
\vy\mid
\widehat{\vxi}_0^{(1)}(t),\ldots,
\widehat{\vxi}_0^{(J)}(t)
\right).
\label{eq:score_guidance_gradient}
\end{equation}
The corresponding guided score,
computed at every reverse time $t$, is
\begin{equation}
\vs_{\mathrm{guided}}^{(j)}(t)
=
\widehat{\vs}^{(j)}(\vxi^{(j)}(t),t)
+\eta^{(j)}(t)\vg^{(j)}(t),
\label{eq:score_guided_score}
\end{equation}
where $\eta^{(j)}(t)$ is the guidance strength.
The guided reverse SDE is
\begin{multline}
d\vxi^{(j)}(t)
=
\left[
\vect{b}_j(\vxi^{(j)}(t),t)
-g_j^2(t)\vs_{\mathrm{guided}}^{(j)}(t)
\right]dt\\
+g_j(t)\,d\overline{\vW}^{(j)}(t),
\label{eq:score_guided_reverse_sde}
\end{multline}
which is integrated backward from $T$ to $0$.

For channel-only estimation, the likelihood
guides the channel reverse state. In blind MIMO
reconstruction, the channel and image reverse
states are both updated at every reverse step
using gradients of the same joint observation
likelihood.

Among $P$ final candidates, we select the one
maximizing the observation likelihood.

Algorithm~\ref{alg:score_guided} summarizes this likelihood-guided
reconstruction procedure.

\begin{algorithm}[t]
\caption{Score-Based Likelihood-Guided Reconstruction}
\label{alg:score_guided}
\begin{algorithmic}[1]
\REQUIRE $\vy$, Gaussian likelihood $p(\vy\mid\cdot)$,
$J$ score-based priors, $\{\eta^{(j)}(t)\}$,
SDE coefficients, reverse grid, $P$

\FOR{$p=1,\ldots,P$}

    \STATE Draw $\{\vxi^{(j,p)}(T)\}_{j=1}^{J}$
    from their terminal marginals

    \FOR{$\ell=N_{\mathrm{grid}},\ldots,1$}

        \STATE $t\gets t_\ell$

        \FOR{$j=1,\ldots,J$}

            \STATE Compute adapted prior score
            $\widehat{\vs}^{(j,p)}(t)$

            \STATE Compute clean estimate
            $\widehat{\vxi}_0^{(j,p)}(t)$
            using \eqref{eq:score_general_clean_estimate}

        \ENDFOR

        \FOR{$j=1,\ldots,J$}

            \STATE Compute likelihood gradient
            $\vg^{(j,p)}(t)$
            using \eqref{eq:score_guidance_gradient}

            \STATE Form guided score
            $\vs_{\mathrm{guided}}^{(j,p)}(t)$
            using \eqref{eq:score_guided_score}

            \STATE Advance $\vxi^{(j,p)}(t)$
            to $\vxi^{(j,p)}(t_{\ell-1})$
            using \eqref{eq:score_guided_reverse_sde}

        \ENDFOR

    \ENDFOR

\ENDFOR

\STATE
$p^\star\gets
\arg\max_{1\leq p\leq P}
p\!\left(
\vy\mid
\vxi^{(1,p)}(0),\ldots,\vxi^{(J,p)}(0)
\right)$

\RETURN
$\{\vxi^{(j,p^\star)}(0)\}_{j=1}^{J}$

\end{algorithmic}
\end{algorithm}

\section{DLPM}
\label{app:dlpm-comprehensive}

\subsection{Forward and Reverse Processes}

Let $\alpha\in(1,2)$ and let $\{\beta_t\}_{t=1}^{T}$ be a noise schedule.
Define
\begin{equation*}
c_t=(1-\beta_t)^{1/\alpha},
\qquad
\sigma_t=\beta_t^{1/\alpha},
\qquad
\bar c_t=\prod_{s=1}^{t}c_s.
\end{equation*}
A DLPM defines the forward process
\begin{equation*}
q(\vX_t\mid\vX_{t-1}=\vx_{t-1})
=
\SaS\!\left(c_t\vx_{t-1},\sigma_t\right).
\end{equation*}
Equivalently,
\begin{equation}
\vX_t
=
\bar c_t\vX_0+s_t\vepsilon_t,
\qquad
s_t=(1-\bar c_t^\alpha)^{1/\alpha},
\qquad
\vepsilon_t\sim\SaS(\vzeros,1).
\label{eq:dlpm_forward_marginal_appendix}
\end{equation}

Using the Gaussian scale-mixture representation in
\eqref{eq:stable_subgaussian_representation}, each perturbation is written as
\begin{equation*}
\vepsilon_t
\overset{\mathrm d}{=}
\sqrt{A_t}\,\vG_t,
\qquad
\vG_t\sim\mathcal{N}(\vzeros,\mat{I}_d),
\end{equation*}
where $A_1,\ldots,A_t$ are independent positive
$\alpha/2$-stable variables with skewness $\beta_A=1$, scale
$\gamma_A=\cos^{2/\alpha}(\pi\alpha/4)$, and location $\delta_A=0$.
Conditioned on
$A_{1:t}=(A_1,\ldots,A_t)$, define
\begin{align}
V_0&=0,
&
V_t&=c_t^2V_{t-1}+\sigma_t^2A_t.
\label{eq:dlpm_conditional_variance}
\end{align}
Then
\begin{equation*}
\vX_t\mid(\vX_0=\vx_0,A_{1:t})
\sim
\mathcal{N}\!\left(
\bar c_t\vx_0,V_t\mat{I}_d
\right).
\end{equation*}
The exact reverse posterior conditioned on $\vX_0=\vx_0$ and $A_{1:t}$ is
\begin{equation*}
q(\vX_{t-1}\mid\vX_t=\vx_t,\vX_0=\vx_0,A_{1:t})
=
\mathcal{N}\!\left(
\widetilde{\vmu}_t,
\widetilde V_t\mat{I}_d
\right),
\end{equation*}
where
\begin{align*}
\widetilde{\vmu}_t
&=
\frac{c_tV_{t-1}}{V_t}\vx_t
+\frac{\bar c_{t-1}\sigma_t^2A_t}{V_t}\vx_0,\\
\widetilde V_t
&=
\frac{\sigma_t^2A_tV_{t-1}}{V_t}.
\end{align*}

The learned reverse transition replaces the unknown marginal perturbation
by $\vepsilon_\theta(\vx_t,t)$. Let
\begin{align*}
\kappa_t&=\frac{\sigma_t^2A_t}{V_t},
\qquad
r_t=\kappa_tV_{t-1}.
\end{align*}
The reverse transition is
\begin{equation*}
p_\theta(\vX_{t-1}\mid\vX_t=\vx_t,A_{1:T})
=
\mathcal{N}\!\left(
\vmu_\theta(\vx_t,t,A_{1:T}),
r_t\mat{I}_d
\right),
\end{equation*}
with
\begin{align*}
\vmu_\theta(\vx_t,t,A_{1:T})
&=
\frac{
\vx_t-\kappa_ts_t\vepsilon_\theta(\vx_t,t)
}{c_t}.
\end{align*}
The corresponding clean estimate is
\begin{equation*}
\widehat{\vx}_0(\vx_t,t)
=
\frac{\vx_t-s_t\vepsilon_\theta(\vx_t,t)}{\bar c_t}.
\end{equation*}
At the beginning of each reverse trajectory, one independent path
$A_{1:T}$ is drawn and \eqref{eq:dlpm_conditional_variance} is evaluated.
Starting from the terminal marginal, reverse sampling uses
\begin{align*}
\vx_{t-1}
&=
\vmu_\theta(\vx_t,t,A_{1:T})
+\sqrt{r_t}\,\widetilde{\vG}_t,
\qquad
\widetilde{\vG}_t\sim\mathcal{N}(\vzeros,\mat{I}_d),
\end{align*}
where $\widetilde{\vG}_1=\vzeros$.

\subsection{Canonical Alpha-Stable Adaptation}
\label{subsec:appendix_dlpm_canonical_adaptation}

We train one DLPM on the canonical distribution
$\vZ_0\sim\SaS(\vzeros,1)$. The DLPM experiments use unit scale
$\gamma=1$; accordingly, the following adaptation changes only the location.
For the unit-scale target distribution
\begin{equation*}
\vX_0=\vmu+\vZ_0 
\sim\SaS(\vmu,1),
\end{equation*}
the forward marginal in
\eqref{eq:dlpm_forward_marginal_appendix} gives
\begin{equation*}
\vX_t
=
\bar c_t\vmu+\vZ_t.
\end{equation*}

\begin{lemma}
Let $\vepsilon_\theta$ be the stable-noise predictor trained on the
canonical distribution. The canonical input and transported prediction are
\begin{equation}
\vU_t=\vX_t-\bar c_t\vmu,
\qquad
\widehat{\vepsilon}_{\vmu,1}(\vX_t,t)
=
\vepsilon_\theta(\vU_t,t).
\label{eq:dlpm_canonical_adaptation_appendix}
\end{equation}
\end{lemma}

\begin{proof}
Substituting $\vX_0=\vmu+\vZ_0$ into
\eqref{eq:dlpm_forward_marginal_appendix} gives
\begin{align*}
\vX_t
&=
\bar c_t(\vmu+\vZ_0)+s_t\overline{\vepsilon}_t\\
&=
\bar c_t\vmu+\vZ_t.
\end{align*}
Therefore, $\vX_t-\bar c_t\vmu$ has the same distribution as the
canonical state $\vZ_t$. Evaluating the canonical predictor at
$\vU_t=\vX_t-\bar c_t\vmu$ proves
\eqref{eq:dlpm_canonical_adaptation_appendix}.
\end{proof}

\subsection{Likelihood-Guided Reconstruction}

We consider the general observation model
\eqref{eq:generic_observation_model}, where each
unknown $\vxi_0^{(j)}$ has a separately trained
diffusion prior. This formulation includes both
channel-only estimation in
\eqref{eq:channel_only_model} and joint MIMO
reconstruction in \eqref{eq:joint_mimo_problem}.

For $\alpha$-stable channel-only estimation, we set
$J=1$, $\vxi_0^{(1)}=\vH_k$, and
$f(\vxi_0^{(1)})=\vxi_0^{(1)}$.
The independent method uses the stationary $\alpha$-stable
channel prior in \eqref{eq:stable_channel_dynamics},
whereas the temporal method uses the predictive prior
in \eqref{eq:stable_predictive_channel_prior}.

For blind joint channel--image MIMO reconstruction, we set
$J=2$,
$\vxi_0^{(1)}=\operatorname{vec}(\mat{H}_k)$,
$\vxi_0^{(2)}=\operatorname{vec}(\mat{D}_k)$, and
$f(\vxi_0^{(1)},\vxi_0^{(2)})
=\operatorname{vec}(\mat{H}_kf_\gamma(\mat{D}_k))$.
The channel and image have separately trained
diffusion priors, coupled through the MIMO likelihood.
The independent method uses the stationary $\alpha$-stable
channel prior in \eqref{eq:stable_channel_dynamics},
whereas the temporal method uses the predictive prior
in \eqref{eq:mimo_stable_temporal_prior}.
The image prior is the same for both methods.

The DLPM experiments use likelihood
guidance. For the observation model in \eqref{eq:generic_observation_model},
\begin{equation*}
\vY
=
f\!\left(\vxi_0^{(1)},\ldots,\vxi_0^{(J)}\right)+\vN,
\qquad
\vN\sim\SaS(\vzeros,\gamma_n),
\end{equation*}
Using the Gaussian scale-mixture representation in \eqref{eq:stable_subgaussian_representation} gives
\begin{equation*}
\vY\mid
\vxi_0^{(1)},\ldots,\vxi_0^{(J)},B
\sim
\mathcal{N}\!\left(
f\!\left(\vxi_0^{(1)},\ldots,\vxi_0^{(J)}\right),
\gamma_n^2B\mat{I}_d
\right),
\end{equation*}
where $B$ is positive $\alpha/2$-stable. Using $M_{\mathrm{MC}}$ independent samples
$B^{(1)},\ldots,B^{(M_{\mathrm{MC}})}$, the likelihood is approximated by
\begin{multline*}
\widehat p\!\left(
\vy\mid\vxi_0^{(1)},\ldots,\vxi_0^{(J)}
\right) = \\
\frac{1}{M_{\mathrm{MC}}}\sum_{m=1}^{M_{\mathrm{MC}}}
\mathcal{N}\!\left(
\vy;
f\!\left(\vxi_0^{(1)},\ldots,\vxi_0^{(J)}\right),
\gamma_n^2B^{(m)}\mat{I}_d
\right).
\end{multline*}
Samples $B^{(m)}$ are used only to approximate the likelihood of measurement
 and are independent of the DLPM trajectory variables
$A_{1:T}$.

For each unknown DLPM and at each time step $t=T,\ldots,1$, the predicted
clean state is
\begin{equation}
\widehat{\vxi}_{0,t}^{(j)}
=
\frac{
\vxi_t^{(j)}-s_t^{(j)}
\widehat{\vepsilon}_t^{(j)}
}{\bar c_t^{(j)}}.
\label{eq:dlpm_general_clean_estimate}
\end{equation}
The observation likelihood gradients are then evaluated as
\begin{equation}
\vg_t^{(j)}
=
\nabla_{\vxi_t^{(j)}}
\log\widehat p\!\left(
\vy\mid
\widehat{\vxi}_{0,t}^{(1)},\ldots,
\widehat{\vxi}_{0,t}^{(J)}
\right).
\label{eq:dlpm_guidance_gradient}
\end{equation}
The corresponding guided reverse update,
performed at every time step, is
\begin{align*}
\vxi_{t-1}^{(j)}
&=
\vmu_{\theta,t}^{(j)}
+\eta_t^{(j)}\vg_t^{(j)}
+\sqrt{r_t^{(j)}}\,\widetilde{\vG}_t^{(j)},
\end{align*}
where $\eta_t^{(j)}$ is the guidance weight and
$\widetilde{\vG}_t^{(j)}\sim\mathcal{N}(\vzeros,\mat{I}_d)$, with
$\widetilde{\vG}_1^{(j)}=\vzeros$.

For channel-only estimation, the likelihood
guides the channel reverse state. 

In blind MIMO
reconstruction, the channel and image reverse
states are both updated at every diffusion step
using gradients of the same joint observation
likelihood. Among $P$ final candidates, we select the one maximizing the approximated observation likelihood.

Algorithm~\ref{alg:dlpm_guided} summarizes this likelihood-guided
reconstruction procedure.

\begin{algorithm}[H]
\caption{DLPM Likelihood-Guided Reconstruction}
\label{alg:dlpm_guided}
\begin{algorithmic}[1]
\REQUIRE $\vy$, approximated likelihood $\widehat p(\vy\mid\cdot)$,
$J$ diffusion priors, $\{\eta_t^{(j)}\}$, diffusion schedules,
$M_{\mathrm{MC}}$, $P$
\STATE Draw the measurement-likelihood samples
$\{B^{(m)}\}_{m=1}^{M_{\mathrm{MC}}}$
\FOR{$p=1,\ldots,P$}
    \STATE Draw an independent $A_{1:T}^{(j,p)}$ for each DLPM prior
    \STATE Compute the corresponding $\{V_t^{(j,p)}\}_{t=1}^{T}$ using
    \eqref{eq:dlpm_conditional_variance}
    \STATE Draw $\{\vxi_T^{(j,p)}\}_{j=1}^{J}$ from their terminal marginals
    \FOR{$t=T,\ldots,1$}
        \FOR{$j=1,\ldots,J$}
            \STATE Compute adapted prediction $\widehat{\vepsilon}_t^{(j,p)}$
            \STATE Compute clean estimate $\widehat{\vxi}_{0,t}^{(j,p)}$
            using \eqref{eq:dlpm_general_clean_estimate} for DLPM
        \ENDFOR
        \FOR{$j=1,\ldots,J$}
            \STATE Compute likelihood gradient $\vg_t^{(j,p)}$ using
            \eqref{eq:dlpm_guidance_gradient}
            \STATE Advance $\vxi_t^{(j,p)}$ to $\vxi_{t-1}^{(j,p)}$
            using the guided reverse update of its prior
        \ENDFOR
    \ENDFOR
\ENDFOR
\IF{channel-only reconstruction}
\STATE $\widehat{\vxi}_0
=
\operatorname{median}_{p=1,\ldots,P}
\left\{
\vxi_0^{(1,p)}
\right\}$
\RETURN $\widehat{\vxi}_0$
\ELSE
\STATE $p^\star\gets
\arg\max_{1\leq p\leq P}
\widehat p\!\left(
\vy\mid
\vxi_0^{(1,p)},\ldots,\vxi_0^{(J,p)}
\right)$
\RETURN
$\left\{\vxi_0^{(j,p^\star)}\right\}_{j=1}^{J}$
\ENDIF
\end{algorithmic}
\end{algorithm}

\section{LIM}
\label{app:lim-comprehensive}

\subsection{Forward and Reverse Processes}

A continuous LIM defines the forward SDE
\begin{align*}
d\vX(t)
&=
\vect{b}(\vX(t),t)\,dt
+g(t)\,d\vL^{\alpha}(t),
\end{align*}
where $\vL^{\alpha}(t)$ is a symmetric $\alpha$-stable L\'{e}vy process.
We consider the linear $\alpha$-stable perturbation kernel associated with
this SDE,
\begin{equation}
\vX(t)
=
a(t)\vX_0+\sigma(t)\vepsilon(t),
\qquad
\vepsilon(t)\sim\SaS(\vzeros,1),
\label{eq:lim_forward_marginal_appendix}
\end{equation}
Let $p_{\vX(t)}$ denote the marginal density of $\vX(t)$. Its fractional
score is
\begin{equation}
\vS_{\vX(t)}^{(\alpha)}(\vx(t))
=
\frac{
(-\Delta)^{\frac{\alpha-2}{2}}\nabla
p_{\vX(t)}(\vx(t))
}{p_{\vX(t)}(\vx(t))}.
\label{eq:lim_fractional_score_appendix}
\end{equation}
The corresponding reverse-time dynamics used by LIM are
\begin{equation}
d\vX(t)
=
\left[
\vect{b}(\vX(t),t)
 -g^{\alpha}(t)\vS_{\vX(t)}^{(\alpha)}(\vX(t))
\right]dt,
\label{eq:lim_reverse_flow_appendix}
\end{equation}
which are solved from $t=T$ to $t=0$. The network
$\vS_\theta^{(\alpha)}(\vX(t),t)$ approximates the fractional score
$\vS_{\vX(t)}^{(\alpha)}(\vX(t))$ and is substituted into
\eqref{eq:lim_reverse_flow_appendix} during generation.

For the perturbation kernel in
\eqref{eq:lim_forward_marginal_appendix}, the clean estimate is
\begin{align*}
\widehat{\vX}_0(\vX(t),t)
&=
\frac{
\vX(t)-\sigma(t)\widehat{\vepsilon}(\vX(t),t)
}{a(t)}.
\end{align*}
Starting from the terminal marginal, the reverse dynamics are evaluated on a
decreasing numerical grid of $N_{\mathrm{grid}}$ intervals,
$T=t_{N_{\mathrm{grid}}}>\cdots>t_0=0$.

\subsection{Canonical Alpha-Stable Adaptation}
\label{subsec:appendix_lim_canonical_adaptation}

We train one LIM on the canonical distribution
$\vZ_0\sim\SaS(\vzeros,1)$ and adapt its fractional score to
$\vX_0\sim\SaS(\vmu,\gamma)$ without retraining. From
\eqref{eq:lim_forward_marginal_appendix},
\begin{align*}
\vX(t)
&=
a(t)\vmu+\gamma\vZ(t),
&
\vU(t)
&=
\frac{\vX(t)-a(t)\vmu}{\gamma}.
\end{align*}
Thus, $\vU(t)$ has the same canonical marginal as $\vZ(t)$.

\begin{lemma}
Let $\vS_\theta^{(\alpha)}$ be the fractional score trained on the canonical
state $\vZ(t)$. The fractional score for
$\vX_0\sim\SaS(\vmu,\gamma)$ is obtained without retraining as
\begin{equation}
\widehat{\vS}_{\vmu,\gamma}^{(\alpha)}(\vX(t),t)
=
\gamma^{1-\alpha}
\vS_\theta^{(\alpha)}(\vU(t),t).
\label{eq:lim_canonical_adaptation_appendix}
\end{equation}
\end{lemma}

\begin{proof}
Let $p_{\vX(t)}$ and $p_{\vU(t)}$ denote the densities of $\vX(t)$ and
$\vU(t)$. From the affine transformation,
\(
\vU(t)
=
\gamma^{-1}\bigl(\vX(t)-a(t)\vmu\bigr).
\)
The density transformation is
\begin{align*}
p_{\vU(t)}(\vU(t))
&=
\gamma^d p_{\vX(t)}\!\left(a(t)\vmu+\gamma\vU(t)\right).
\end{align*}
Using
$\widehat q(\vect{f})=\int q(\vX(t))
e^{-i2\pi\vect{f}^{\mathsf T}\vX(t)}d\vX(t)$,
\begin{align*}
\widehat p_{\vU(t)}(\vect{f})
&=
e^{i2\pi\vect{f}^{\mathsf T}a(t)\vmu/\gamma}
\widehat p_{\vX(t)}\!\left(\frac{\vect{f}}{\gamma}\right).
\end{align*}
By the fractional-score definition in
\eqref{eq:lim_fractional_score_appendix},
\begin{align*}
\mathcal{F}\!\left\{
\vS_{\vX(t)}^{(\alpha)}p_{\vX(t)}
\right\}(\vect{f})
&=
i2\pi\vect{f}(2\pi\|\vect{f}\|_2)^{\alpha-2}
\widehat p_{\vX(t)}(\vect{f}).
\end{align*}
\(\Longrightarrow\quad
\begin{aligned}[t]
&\mathcal{F}\!\left\{
\vS_{\vX(t)}^{(\alpha)}p_{\vX(t)}
\right\}\!\left(\frac{\vect{f}}{\gamma}\right)
\\
&=
i2\pi\frac{\vect{f}}{\gamma}
\left(
2\pi\left\lVert\frac{\vect{f}}{\gamma}\right\rVert_2
\right)^{\alpha-2}
\widehat p_{\vX(t)}
\!\left(\frac{\vect{f}}{\gamma}\right)
\\
&=
\gamma^{1-\alpha}
i2\pi\vect{f}
\left(2\pi\lVert\vect{f}\rVert_2\right)^{\alpha-2}
\widehat p_{\vX(t)}
\!\left(\frac{\vect{f}}{\gamma}\right).
\end{aligned}
\)

Therefore,
\begin{align*}
&\mathcal{F}\!\left\{
\vS_{\vU(t)}^{(\alpha)}p_{\vU(t)}
\right\}(\vect{f})\\
&=
i2\pi\vect{f}(2\pi\|\vect{f}\|_2)^{\alpha-2}
\widehat p_{\vU(t)}(\vect{f})\\
&=
\gamma^{\alpha-1}
e^{i2\pi\vect{f}^{\mathsf T}a(t)\vmu/\gamma}
\mathcal{F}\!\left\{
\vS_{\vX(t)}^{(\alpha)}p_{\vX(t)}
\right\}\!\left(\frac{\vect{f}}{\gamma}\right).
\end{align*}
The Fourier scaling-and-translation property also gives
\begin{align*}
&\mathcal{F}_{\vU(t)}\!\left\{
\vS_{\vX(t)}^{(\alpha)}\!\left(
a(t)\vmu+\gamma\vU(t)
\right)
p_{\vX(t)}\!\left(
a(t)\vmu+\gamma\vU(t)
\right)
\right\}(\vect{f})\\
&=
\gamma^{-d}
e^{i2\pi\vect{f}^{\mathsf T}a(t)\vmu/\gamma}
\mathcal{F}_{\vX(t)}\!\left\{
\vS_{\vX(t)}^{(\alpha)}p_{\vX(t)}
\right\}\!\left(\frac{\vect{f}}{\gamma}\right).
\end{align*}
Hence,
\begin{align*}
&\mathcal{F}_{\vU(t)}\!\left\{
\vS_{\vU(t)}^{(\alpha)}p_{\vU(t)}
\right\}(\vect{f})\\
&=
\gamma^{d+\alpha-1}
\mathcal{F}_{\vU(t)}\!\left\{
\vS_{\vX(t)}^{(\alpha)}\!\left(
a(t)\vmu+\gamma\vU(t)
\right)
p_{\vX(t)}\!\left(
a(t)\vmu+\gamma\vU(t)
\right)
\right\}(\vect{f}).
\end{align*}
Taking the inverse Fourier transform gives
\begin{align*}
&\vS_{\vU(t)}^{(\alpha)}(\vU(t))
p_{\vU(t)}(\vU(t))\\
&=
\gamma^{d+\alpha-1}
\vS_{\vX(t)}^{(\alpha)}\!\left(
a(t)\vmu+\gamma\vU(t)
\right)
p_{\vX(t)}\!\left(a(t)\vmu+\gamma\vU(t)\right).
\end{align*}
Dividing by
$p_{\vU(t)}(\vU(t))
=\gamma^d p_{\vX(t)}(a(t)\vmu+\gamma\vU(t))$ gives
\begin{align*}
\vS_{\vU(t)}^{(\alpha)}(\vU(t))
&=
\gamma^{\alpha-1}
\vS_{\vX(t)}^{(\alpha)}(\vX(t)),
\end{align*}
and hence
\begin{align*}
\vS_{\vX(t)}^{(\alpha)}(\vX(t))
&=
\gamma^{1-\alpha}
\vS_{\vU(t)}^{(\alpha)}(\vU(t))
\end{align*}
establishes
\eqref{eq:lim_canonical_adaptation_appendix}.
\end{proof}

\subsection{Likelihood-Guided Reconstruction}

We consider the general observation model
\eqref{eq:generic_observation_model}, where each unknown
$\vxi_0^{(j)}$ has a separately trained diffusion prior. This formulation
includes both channel-only estimation in \eqref{eq:channel_only_model} and
joint MIMO reconstruction in \eqref{eq:joint_mimo_problem}.

For $\alpha$-stable channel-only estimation, we set
$J=1$, $\vxi_0^{(1)}=\vH_k$, and
$f(\vxi_0^{(1)})=\vxi_0^{(1)}$. The independent method uses the stationary
$\alpha$-stable channel prior in \eqref{eq:stable_channel_dynamics}, whereas
the temporal method uses the predictive prior in
\eqref{eq:stable_predictive_channel_prior}.

For blind joint channel--image MIMO reconstruction, we set
$J=2$, $\vxi_0^{(1)}=\operatorname{vec}(\mat{H}_k)$,
$\vxi_0^{(2)}=\operatorname{vec}(\mat{D}_k)$, and
$f(\vxi_0^{(1)},\vxi_0^{(2)})
=\operatorname{vec}(\mat{H}_kf_\gamma(\mat{D}_k))$.
The channel and image have separately trained diffusion priors, coupled
through the MIMO likelihood. The independent method uses the stationary
$\alpha$-stable channel prior in \eqref{eq:stable_channel_dynamics}, whereas
the temporal method uses the predictive prior in
\eqref{eq:mimo_stable_temporal_prior}. The image prior is the same for both
methods.

The LIM experiments use likelihood guidance. For the observation model in \eqref{eq:generic_observation_model},
\begin{equation*}
\vY
=
f\!\left(\vxi_0^{(1)},\ldots,\vxi_0^{(J)}\right)+\vN,
\qquad
\vN\sim\SaS(\vzeros,\gamma_n),
\end{equation*}
Using the Gaussian scale-mixture representation in \eqref{eq:stable_subgaussian_representation} gives
\begin{equation*}
\vY\mid
\vxi_0^{(1)},\ldots,\vxi_0^{(J)},B
\sim
\mathcal{N}\!\left(
f\!\left(\vxi_0^{(1)},\ldots,\vxi_0^{(J)}\right),
\gamma_n^2B\mat{I}_d
\right),
\end{equation*}
where $B$ is positive $\alpha/2$-stable. Using $M_{\mathrm{MC}}$ independent samples
$B^{(1)},\ldots,B^{(M_{\mathrm{MC}})}$, the likelihood is approximated by
\begin{multline*}
\widehat p\!\left(
\vy\mid\vxi_0^{(1)},\ldots,\vxi_0^{(J)}
\right)=\\
\frac{1}{M_{\mathrm{MC}}}\sum_{m=1}^{M_{\mathrm{MC}}}
\mathcal{N}\!\left(
\vy;
f\!\left(\vxi_0^{(1)},\ldots,\vxi_0^{(J)}\right),
\gamma_n^2B^{(m)}\mat{I}_d
\right).
\end{multline*}

For each LIM unknown and at every reverse time $t$, the predicted clean state
is
\begin{equation}
\widehat{\vxi}_0^{(j)}(t)
=
\frac{
\vxi^{(j)}(t)-\sigma_j(t)
\widehat{\vepsilon}^{(j)}(\vxi^{(j)}(t),t)
}{a_j(t)}.
\label{eq:lim_general_clean_estimate}
\end{equation}
The observation likelihood gradients are then evaluated as
\begin{equation}
\vg^{(j)}(t)
=
\nabla_{\vxi^{(j)}(t)}
\log\widehat p\!\left(
\vy\mid
\widehat{\vxi}_0^{(1)}(t),\ldots,
\widehat{\vxi}_0^{(J)}(t)
\right).
\label{eq:lim_guidance_gradient}
\end{equation}
At every reverse time, the prior probability-flow step from $t_\ell$ to
$t_{\ell-1}$ is first evaluated using
\eqref{eq:lim_reverse_flow_appendix}. The likelihood correction is then
applied as
\begin{equation}
\vxi^{(j)}(t_{\ell-1})
=
\vxi_{\mathrm{prior}}^{(j)}(t_{\ell-1})
+\eta^{(j)}(t_\ell)
|t_\ell-t_{\ell-1}|\vg^{(j)}(t_\ell),
\label{eq:lim_guided_update}
\end{equation}
where $\eta^{(j)}(t)$ is the guidance strength.

\begin{algorithm}[ht!]
\caption{LIM Likelihood-Guided Reconstruction}
\label{alg:lim_guided}
\begin{algorithmic}[1]
\REQUIRE $\vy$, approximated likelihood $\widehat p(\vy\mid\cdot)$,
$J$ diffusion priors, $\{\eta^{(j)}(t)\}$, reverse grid,
$M_{\mathrm{MC}}$, $P$
\STATE Draw the measurement-likelihood samples
$\{B^{(m)}\}_{m=1}^{M_{\mathrm{MC}}}$
\FOR{$p=1,\ldots,P$}
    \STATE Draw $\{\vxi^{(j,p)}(T)\}_{j=1}^{J}$ from their terminal marginals
    \FOR{$\ell=N_{\mathrm{grid}},\ldots,1$}
        \STATE $t\gets t_\ell$
        \FOR{$j=1,\ldots,J$}
            \STATE Compute adapted fractional score and perturbation estimate
            \STATE Compute clean estimate $\widehat{\vxi}_0^{(j,p)}(t)$
            using \eqref{eq:lim_general_clean_estimate}
        \ENDFOR
        \FOR{$j=1,\ldots,J$}
            \STATE Compute likelihood gradient $\vg^{(j,p)}(t)$ using
            \eqref{eq:lim_guidance_gradient}
            \STATE Advance $\vxi^{(j,p)}(t)$ to
            $\vxi^{(j,p)}(t_{\ell-1})$ using
            \eqref{eq:lim_guided_update}
        \ENDFOR
    \ENDFOR
\ENDFOR
\IF{channel-only reconstruction}
\STATE $\widehat{\vxi}_0
=
\operatorname{median}_{p=1,\ldots,P}
\left\{
\vxi_0^{(1,p)}
\right\}$
\RETURN $\widehat{\vxi}_0$
\ELSE
\STATE $p^\star\gets
\arg\max_{1\leq p\leq P}
\widehat p\!\left(
\vy\mid
\vxi_0^{(1,p)},\ldots,\vxi_0^{(J,p)}
\right)$
\RETURN
$\left\{\vxi_0^{(j,p^\star)}\right\}_{j=1}^{J}$
\ENDIF
\end{algorithmic}
\end{algorithm}

For channel-only estimation, the likelihood guides the channel reverse
state. In blind MIMO reconstruction, the channel and image reverse states
are both updated at every reverse step using gradients of the same joint
observation likelihood. Among $P$ final candidates, we select the one
maximizing the approximated observation likelihood. 

Algorithm~\ref{alg:lim_guided} summarizes this likelihood-guided
reconstruction procedure.
\section{Additional Continuous-Time Results}
\label{app:continuous-results}

\end{document}